\documentclass{article}

\usepackage{cancel}
\usepackage{tikz}

\usepackage{pgfplots}
\pgfplotsset{compat=1.5}

\newcommand{\TQBF}{\texttt{TQBF}}
\newcommand{\DD}{\texttt{D-ATDP}}
\newcommand{\DO}{\texttt{D-ATOP}}
\newcommand{\ND}{\texttt{N-ATDP}}
\newcommand{\NO}{\texttt{N-ATOP}}
\newcommand{\MSC}{\texttt{MSC}}

\newcommand{\PSPACE}{\texttt{PSPACE}}
\newcommand{\PSPACEc}{\texttt{PSPACE-complete}}
\newcommand{\PSPACEh}{\texttt{PSPACE-hard}}

\newcommand{\LogAPX}{\texttt{Log-APX}}
\newcommand{\LogAPXc}{\texttt{Log-APX-complete}}

\newcommand{\LogAPXh}{\texttt{Log-APX-hard}}

\pgfplotsset{soldot/.style={color=blue,only marks,mark=*}} \pgfplotsset{holdot/.style={color=blue,fill=white,only marks,mark=*}}

\usepackage{mdframed}
\newdimen\arrowsize
\pgfarrowsdeclare{squarea}{squarea}
{
  \arrowsize=0.4pt
  \advance\arrowsize by.275\pgflinewidth%
  \pgfarrowsleftextend{+-\arrowsize}
  \advance\arrowsize by.5\pgflinewidth
  \pgfarrowsrightextend{+\arrowsize}
}
{
  \arrowsize=0.4pt
  \advance\arrowsize by.275\pgflinewidth%
  \pgfsetdash{}{+0pt}
  \pgfsetroundjoin
  \pgfpathmoveto{\pgfqpoint{1\arrowsize}{4\arrowsize}}
  \pgfpathlineto{\pgfqpoint{-7\arrowsize}{4\arrowsize}}
  \pgfpathlineto{\pgfqpoint{-7\arrowsize}{-4\arrowsize}}
  \pgfpathlineto{\pgfqpoint{1\arrowsize}{-4\arrowsize}}
  \pgfpathclose
  \pgfusepathqfillstroke
}

\pgfarrowsdeclare{open squarea}{open squarea}
{
  \arrowsize=0.4pt
  \advance\arrowsize by.275\pgflinewidth%
  \pgfarrowsleftextend{+-.5\pgflinewidth}
  \advance\arrowsize by7\arrowsize
  \advance\arrowsize by.5\pgflinewidth
  \pgfarrowsrightextend{+\arrowsize}
}
{
  \arrowsize=0.4pt
  \advance\arrowsize by.275\pgflinewidth%
  \pgfsetdash{}{+0pt}
  \pgfsetroundjoin
  \pgfpathmoveto{\pgfqpoint{8\arrowsize}{4\arrowsize}}
  \pgfpathlineto{\pgfqpoint{0\arrowsize}{4\arrowsize}}
  \pgfpathlineto{\pgfqpoint{0\arrowsize}{-4\arrowsize}}
  \pgfpathlineto{\pgfqpoint{8\arrowsize}{-4\arrowsize}}
  \pgfpathclose
  \pgfusepathqstroke
}

\makeatletter
\newdimen\tempa
\newdimen\tempb
\pgfdeclareshape{diamond in circle}{
\inheritsavedanchors[from=diamond]
\inheritsavedanchors[from=circle]
\inheritanchorborder[from=circle]
\inheritanchor[from=circle]{center}
\inheritanchor[from=circle]{radius}
\inheritanchor[from=circle]{north}
\inheritanchor[from=circle]{south}
\inheritanchor[from=circle]{east}
\inheritanchor[from=circle]{west}
\inheritanchor[from=circle]{anchorborder}
  \saveddimen\radius{%
    \pgf@ya=.5\ht\pgfnodeparttextbox%
    \advance\pgf@ya by.5\dp\pgfnodeparttextbox%
    \pgfmathsetlength\pgf@yb{\pgfkeysvalueof{/pgf/inner ysep}}%
    \advance\pgf@ya by\pgf@yb%
    \pgf@xa=.5\wd\pgfnodeparttextbox%
    \pgfmathsetlength\pgf@xb{\pgfkeysvalueof{/pgf/inner xsep}}%
    \advance\pgf@xa by\pgf@xb%
    \pgf@process{\pgfpointnormalised{\pgfqpoint{\pgf@xa}{\pgf@ya}}}%
    \ifdim\pgf@x>\pgf@y%
        \c@pgf@counta=\pgf@x%
        \ifnum\c@pgf@counta=0\relax%
        \else%
          \divide\c@pgf@counta by 255\relax%
          \pgf@xa=16\pgf@xa\relax%
          \divide\pgf@xa by\c@pgf@counta%
          \pgf@xa=16\pgf@xa\relax%
        \fi%
      \else%
        \c@pgf@counta=\pgf@y%
        \ifnum\c@pgf@counta=0\relax%
        \else%
          \divide\c@pgf@counta by 255\relax%
          \pgf@ya=16\pgf@ya\relax%
          \divide\pgf@ya by\c@pgf@counta%
          \pgf@xa=16\pgf@ya\relax%
        \fi%
    \fi%
    \pgf@x=\pgf@xa%
    \pgfmathsetlength{\pgf@xb}{\pgfkeysvalueof{/pgf/minimum width}}%
    \pgfmathsetlength{\pgf@yb}{\pgfkeysvalueof{/pgf/minimum height}}%
    \ifdim\pgf@x<.5\pgf@xb%
        \pgf@x=.5\pgf@xb%
    \fi%
    \ifdim\pgf@x<.5\pgf@yb%
        \pgf@x=.5\pgf@yb%
    \fi%
    \pgfmathsetlength{\pgf@xb}{\pgfkeysvalueof{/pgf/outer xsep}}%
    \pgfmathsetlength{\pgf@yb}{\pgfkeysvalueof{/pgf/outer ysep}}%
    \ifdim\pgf@xb<\pgf@yb%
      \advance\pgf@x by\pgf@yb%
    \else%
      \advance\pgf@x by\pgf@xb%
    \fi%
  }
\backgroundpath{
    \tempa=\radius
    \pgfmathsetlength{\pgf@xb}{\pgfkeysvalueof{/pgf/outer xsep}}%
    \pgfmathsetlength{\pgf@yb}{\pgfkeysvalueof{/pgf/outer ysep}}%
    \ifdim\pgf@xb<\pgf@yb%
      \advance\tempa by-\pgf@yb%
    \else%
      \advance\tempa by-\pgf@xb%
    \fi%
    \pgfpathmoveto{\centerpoint\advance\pgf@x by\radius}%
    \pgfpathlineto{\centerpoint\advance\pgf@y by\radius}%
    \pgfpathlineto{\centerpoint\advance\pgf@x by-\radius}%
    \pgfpathlineto{\centerpoint\advance\pgf@y by-\radius}%
    \pgfpathclose%
  }
\behindbackgroundpath{
    \tempa=\radius%
    \pgfmathsetlength{\pgf@xb}{\pgfkeysvalueof{/pgf/outer xsep}}%
    \pgfmathsetlength{\pgf@yb}{\pgfkeysvalueof{/pgf/outer ysep}}%
    \ifdim\pgf@xb<\pgf@yb%
      \advance\tempa by-\pgf@yb%
    \else%
      \advance\tempa by-\pgf@xb%
    \fi%
    \pgfpathcircle{\centerpoint}{\tempa}%
  }
}
\makeatother

\usepackage{eurosym}

\usepackage[english]{babel}
\usepackage{amsfonts}
\usepackage{amsmath}
\usepackage{mathtools}
\usepackage{amssymb}
\usepackage{epsfig}
\usepackage{alltt}
\usepackage[latin1]{inputenc}
\usepackage{fixme}

\usepackage{enumerate}
\usepackage[T1]{fontenc}
\usepackage[latin1]{inputenc}
\usepackage{amssymb,amsmath}
\usepackage{hyperref}

\newcounter{cdobleimpl}
\def\thecdobleimpl{\ifnum\value{cdobleimpl}=1 $\Longrightarrow$:\ \else $\Longleftarrow$:\ \fi}

\newenvironment{proof}{\par\textbf{\textit{Proof.}}\ }{\qed}

\def\squareforqed{\hbox{\rlap{$\sqcap$}$\sqcup$}}
\def\qed{\ifmmode\squareforqed\else{\unskip\nobreak\hfil
\penalty50\hskip1em\null\nobreak\hfil\squareforqed
\parfillskip=0pt\finalhyphendemerits=0\endgraf}\fi}

\newtheorem{theorem}{\bfseries Theorem}
\newtheorem{definition}{\bfseries Definition}
\newtheorem{proposition}{\bfseries Proposition}
\newtheorem{corollary}{\bfseries Corollary}
\newtheorem{lemma}{\bfseries Lemma}
\newtheorem{example}{\bfseries Example}

\newcommand{\ismcomment}[1]{}
\newcommand{\bdfn}{\begin{definition} \begin{rm}}
\newcommand{\edfn}{\end{rm}$ $\qed \end{definition}}
\newcommand{\bthm}{\begin{theorem} \begin{rm}}
\newcommand{\ethm}{\end{rm}$ $\qed \end{theorem}}
\newcommand{\bprop}{\begin{proposition} \begin{rm}}
\newcommand{\eprop}{\end{rm}\qed\end{proposition}}
\newcommand{\bcor}{\begin{corollary}\begin{rm}}
\newcommand{\ecor}{\end{rm} \end{corollary}}
\newcommand{\blem}{\begin{lemma} \begin{rm}}
\newcommand{\elem}{\end{rm}\qed\end{lemma}}
\newcommand{\bfact}{\begin{fact} \begin{rm}}
\newcommand{\efact}{\end{rm} \end{fact}}
\newcommand{\bex}{\begin{example} \begin{rm}}
\newcommand{\eex}{\end{rm}$ $\qed  \end{example}}

\newcommand{\bprf}{\begin{proof}}
\newcommand{\eprf}{\end{proof}}

\newenvironment{sketch}%
{\nopagebreak[4]\vspace*{0.2em}\noindent{\bf\it Proof Sketch}:\hspace{1ex}}
  {}
\newcommand{\bprfsketch}{\begin{sketch}}
\newcommand{\eprfsketch}{\end{sketch}}

\newcommand{\comen}[1]{}

\def\y{\;\wedge\;}
\def\o{\;\vee\;}

\newcommand{\letbar}[1]{\mbox{I\kern-0.23em#1}}
\newcommand{\nat}{\mathbb{N}}

\newcommand{\NP}{\texttt{NP}}
\newcommand{\NPc}{\texttt{NP-complete}}
\newcommand{\NPh}{\texttt{NP-hard}}

\newcommand{\calC}{{\cal C}}

\newcommand{\calI}{{\cal I}}

\newcommand{\calP}{{\cal P}}

\newcommand{\si}{\mathtt{if\  }}

\newcommand{\eoc}{\mathtt{otherwise\  }}

\newbox\arriba
\newbox\abajo
\newbox\CaracterInterno
\newbox\CaracterDerecha
\newdimen\anchura
\def\MacrosTranGeneral#1#2#3#4#5#6{%
  \setbox\CaracterInterno=\hbox{\mathsurround=0pt$\mathord#4$}
  \setbox\CaracterDerecha=\hbox{\mathsurround=0pt$\mathord#3$}
  \setbox\arriba=\hbox{$#1#2$}
  \setbox\abajo=\hbox{\mathsurround=0pt%
                      \anchura=\wd\arriba%
                      \advance \anchura by 0.5em%
                      \divide \anchura by \wd\CaracterInterno%
                      \multiply \anchura by \wd\CaracterInterno%
                      \copy\CaracterInterno\kern\SeparacionInternaFlecha
                      \hbox to \anchura{%
                          $\cleaders%
                            \hbox{\kern\SeparacionInternaFlecha\copy\CaracterInterno}
                            \hfill$}%
                      \kern\SeparacionExternaFlecha\copy\CaracterDerecha}
  \mathrel{{\buildrel\vbox{\copy\arriba \kern\SeparacionFlechaArriba} %
    \over{\copy\abajo^{#6}}}_{#5}}
  }
\def\MacrosTranGeneralProp#1#2#3#4#5{\mathchoice%
  {\MacrosTranGeneral{\scriptstyle}{#1}{#2}{#3}{#4}{#5}}
  {\MacrosTranGeneral{\scriptstyle}{#1}{#2}{#3}{#4}{#5}}
  {\MacrosTranGeneral{\scriptscriptstyle}{#1}{#2}{#3}{#4}{#5}}
  {\MacrosTranGeneral{\scriptscriptstyle}{#1}{#2}{#3}{#4}{#5}}}

\def\MacrosNoTran#1{%
  \def\SeparacionInternaFlecha{-0.3em}
  \def\SeparacionExternaFlecha{-0.5em}
  \def\SeparacionFlechaArriba{-3pt}
  \MacrosTranGeneralProp{#1\kern 0.5em}{{\not\rightarrow}}{-}{}{}}

\renewcommand{\emptyset}{\varnothing}

\title{Complexity of adaptive testing in scenarios defined extensionally\thanks{This paper was published in Frontiers of Computer Science.
The present version is the author's accepted manuscript.}}

\author{Ismael Rodr{\'\i}guez$^{1,3}$, David Rubio$^{2}$,
        and Fernando Rubio$^{1,3}$
\thanks{$^{1}$Dpto. Sistemas Inform{\'a}ticos y Computaci{\'o}n, Facultad de Inform{\'a}tica, Universidad Complutense de Madrid, 28040 Madrid, Spain.}
\thanks{$^{2}$Instituto de Biomec\'anica de Valencia. Universitat Polit\`ecnica de Val\`encia, 46022 Val\`encia, Spain.}
\thanks{$^{3}$Instituto de Tecnolog{\'\i}as del Conocimiento, Universidad Complutense de Madrid, 28040 Madrid, Spain.}
}

\begin{document}

\date{}

\maketitle

\begin{abstract}
In this paper we consider a testing setting where the set of possible definitions of the Implementation Under Test (IUT), as well as the behavior of each of these definitions in all possible interactions, are extensionally defined, i.e., on an element-by-element and case-by-case basis. Under this setting, the problem of finding the minimum testing strategy such that collected observations will necessarily let us decide whether the IUT is correct or not (i.e., whether it necessarily belongs to the set of possible correct definitions or not) is studied in four possible problem variants: with or without non-determinism; and with or without more than one possible definition in the sets of possible correct and incorrect definitions. The computational complexity of these variants is studied, and properties such as  PSPACE-completeness and Log-APX-hardness are identified.
\end{abstract}

PSPACE-completeness, approximation hardness.

\section{Introduction}\label{sec:introduction}

The research on Formal Testing Techniques (FTT) has provided lots of models to reason about testing in many possible different settings. Many aspects affecting the testing activity can be considered in formal testing models: The language the implementation under test (IUT) is assumed to be written in can be, for instance, deterministic or non-deterministic finite state machines, i.e., FSMs~\cite{ly96,pet00,DBLP:journals/infsof/DorofeevaEMCY10}, input/output labelled transition systems~\cite{tre96,tre99,bt00}, temporal automata~\cite{sva01,kt04,bcf06,mrr08}, probabilistic systems~\cite{sv03,lnr06,efatmaneshnik19}, etc.; the set of divergences the IUT is assumed to be able to expose with respect to the desired behavior can be given by e.g., a fault model~\cite{morell1990theory}, a set of hypotheses about the IUT~\cite{hie02b,hie09,rmn08}, etc.; the way we interact with the IUT or collect its responses can be e.g., via delayed interactions, imprecise observations~\cite{rlr14,rrr20}, etc.

A particularly interesting aspect of the latter kind is whether the test cases to be applied to the IUT must be decided before the testing activity begins, called {\it preset} testing, or whether the next inputs to be applied may dynamically depend on the observations (outputs) collected during the application of the previous inputs or test cases, called {\it adaptive} testing~\cite{ly96,ky15,yyk17,kyy16,py11,py14}. In the latter case, lesser testing effort could provide more information on the  \hbox{(in-)correctness} of the IUT, as previous observations can be used to make  subsequent interactions guide the IUT into states providing more  sensitive information  regarding the IUT  (in-)correctness.

Efforts have been made to develop strategies of adaptive testing for different frameworks. As usual, these strategies depend on the language the IUT is assumed to be defined in. For instance, the study of adaptive testing of FSMs has been particularly active~\cite{ly96,ky15}, although recently labelled transition systems have been considered as well~\cite{bv19}, and adaptive testing strategies have also been derived from temporal logic specifications~\cite{bfg19}.
Approaches to adaptive testing also depend on the goal the tester aims at, e.g., assuring conformance testing~\cite{py11,py14}, constructing homing or synchronization sequences~\cite{ky15,kyy16,yyk17}, etc.
Despite these differences, the  challenge of designing adaptive strategies is essentially the same in all approaches if it is seen with enough abstraction: we need to dynamically choose the next moves of the tester in such a way that the significance of the information extracted by subsequent observations is improved. In fact, typically adaptive testing can be seen as a {\it game} between the tester and the IUT, where the tester aims at exposing IUT faults and the IUT aims at hiding them. Certainly the IUT does not {\it aim} at anything, but a strategy discovering the IUT faults {\it in any case} requires covering also the cases where the IUT (unintentionally) hides its faults particularly well, so reasoning about these worst cases is {\it as if} the IUT could decide what to show, within its limited freedom, to make things harder to the tester.
Could this goal, shared by all adaptive testing settings, be defined and studied in a {\it general} way, i.e., not being dependent on the particular IUT language, the particular form of its faults, or any other particular difficulties of the testing activity?

During the last years, some pieces of work have provided models to reason about testing from very general and abstract points of view, regardless of the language the IUT is defined in, the assumed fault model or hypotheses, or the particular conformance criterion. They include general models of {\it testability}, measured in terms of the cardinality required by test suites to be complete~\cite{rod09c,rlr14}, in particular whether they are e.g., finite, countable infinite, inexistent, etc., and they also include {\it testing complexity}, measured in terms of the speed our certainty on the IUT correctness increases with the number of applied test cases~\cite{rrr20}.
Despite their high generality, these models do not regard testing as a process where subsequent actions depend on previous observations, which is the essence of adaptive testing.
In this paper we study the computational complexity of adaptive testing with a simple model designed to capture its essence at its most basic level. %
In fact, instead of providing a model with enough generality to fully express any previous model on adaptive testing, our way to grasp the essential difficulty of adaptive testing {\it in general} will be quite the opposite, as we will study its most basic, simple, essential, and {\it less} compact form: the case where the set of possible behaviors of the IUT for all possible interactions is extensionally defined.

This {\it extensional definition} means we assume the actual definition of the (black box) IUT belongs to a given finite set of possible IUT definitions which are described {\it one by one}. Thus, this view is opposite to compactly characterizing the set of possible definitions of the IUT as all definitions fulfilling some given rule, such as e.g., ``{\it the IUT belongs to the set of all FSMs defined like some given specification FSM but with at most one output fault}'' or ``{\it the IUT is any Java program with up to 1,000 code lines}.''
Moreover, our extensional definition view consistently applies to the definition of each possible IUT definition itself,
as the description of all possible ways of interaction with each possible IUT definition is also given {\it one by one}. Hence, the set of these interactions is assumed to be finite and, consequently, the sets of possible inputs and outputs to be sent to or received from the IUT, respectively, are finite as well.
All input applications to the IUT are assumed to be independent to each other, so it is assumed that the IUT reliably returns to its initial configuration after each input. Thus, inputs can be regarded as test cases, and outputs as their observed results.\footnote{This means the application of a {\it sequence} of consecutive machine inputs $i_1 \ldots i_n$ to the IUT must be abstracted in this model by a {\it single input} $\sigma$ representing the whole interaction sequence $i_1 \ldots i_n$. Since finite sets of inputs and outputs are assumed, a finite set of possible {\it interaction sequences} $\sigma_1, \ldots, \sigma_m$ with the IUT would be considered in any case.}

Therefore, we extensionally describe what the IUT can do in all situations: we list all its possible complete behavior definitions, where each of them lists all its possible (generally non-deterministic) responses (i.e., outputs) for all possible inputs.
In addition, an extensionally defined subset of the set of all possible definitions of the IUT will denote those considered {\it correct} according to the requirements of the tester.
The goal of the tester will be finding out an adaptive testing strategy  determining the \hbox{(in-)correctness} of the IUT \hbox{---and} if it exists, preferably the shortest one. Deciding whether the IUT is correct with an adaptive strategy means applying some input to the IUT such that, for any output responded by the IUT, the tester can react applying some input next --in general, different for each previous output-- such that, for each output responded by the IUT, the tester can give some input afterwards, and so on in such a way that, in all cases, the tester can eventually classify the IUT as necessarily correct, i.e., inside the subset mentioned before, or necessarily incorrect (outside).

Despite the apparent excessive specificity of the previous model, it is particularly interesting for two reasons. On the one hand, note that it is much more interesting finding lower bounds of the computational complexity of the most {\it particular and simple} versions of problems, as proving the hardness of a problem A on some complexity class trivially implies the hardness on that class of any problem B {\it generalizing} problem A. Thus, the hardness of adaptive testing for any model trivially allowing us to express our extensionally defined model (e.g., the more expressive FSM model can easily be used to denote our extensional model) will be, at least, that found for our extensional model. We will prove the complexity hardness of some variants of adaptive testing for extensionally defined settings in classes such as PSPACE and Log-APX. Let us recall that input applications are assumed to be independent from each other in this model, so these hardness results are actually obtained for a {\it memory-less} model. This shows something that could be surprising at a first glance: the states inherent to any computation model are {\it not} needed at all to make adaptive testing a hard task, even a PSPACE-hard one.

On the other hand, note that there are many software engineering situations where the possible test cases the tester could apply to the IUT are extensionally defined: instead of automatically extracting test cases from a given expressive specification model and perhaps a formally defined fault model, which constitutes a kind of ideal testing methodology, testers just manually define a finite set of test cases and the possible responses of the IUT for each one, where each wrong response denotes a possible fault at the IUT. Since IUT responses for interactions not covered by these manually defined test cases will not be tested in any case, and thus will not provide any information to the tester, this model is equivalent to listing, element by element, a finite set of possible faulty definitions of the IUT, where for each one we define the finite set of possible responses for each possible test case. This pragmatic and trivial model or, roughly speaking, lack of model, which is ultimately very used by testing practitioners, turns out to be equivalent to our extensionally defined model.
Indeed, for many practitioners the specification is just the set of all possible definitions that would respect all the requirements, and the fault model is just the set of all possible wrong definitions the system could have according to the mistakes they explicitly consider as feasible. In these cases the nature of both is given, by definition, by extensionally defined sets: everything in the lists is possible, and everything outside is not ---and any possible analogy between test data is exactly that emerging among the regularities of these definitions.

The computational complexity of adaptive testing under our extensional model will be studied for four different problem variants, resulting from combining two binary choices. On the one hand, we will consider that non-determinism can be either allowed or forbidden.
On the other hand, we will consider the case that either the subset of possible correct definitions of the IUT or the subset of incorrect ones are singletons, or the case that both sets have more than one element.\footnote{We assume the testing process stops as soon as the observations let us either discard all possible correct IUT definitions or discard all possible wrong IUT definitions ---regardless of whether there are still more than one possible definition of the other kind. Note this condition is sufficient to provide a precise incorrectness/correctness diagnosis, respectively. We do not aim at {\it uniquely} identifying the IUT definition within the set of possibilities, which would require a stronger condition.}  The computational complexity of the four resulting problems will be identified, yielding complexity hardness in different classes: both cases requiring determinism are NP-complete, one being Log-APX-complete and the other one being hard, at least, in Log-APX; both cases dealing with non-determinism are PSPACE-complete. Note that no polynomial-time algorithm can find exact solutions for the former problems unless P $=$ NP. Moreover, given the hardness of these problems in Log-APX, it is known that no polynomial-time approximation algorithm can guarantee any constant {\it performance ratio} unless P $=$ NP, that is, the ratio between the quality of returned solutions and the quality of optimal solutions can not be guaranteed bo be bounded by a constant. Regarding the latter problems, i.e. those where non-determinism is permitted, their hardness in PSPACE means no polynomial time algorithm can solve them unless P $=$ PSPACE, which is a stronger and even less expected property than P $=$ NP, given P $\subseteq$ NP $\subseteq$ PSPACE. These complexity hardness results have other practical consequences: our NP-complete problems can be approached by fast sub-optimal metaheuristics such as e.g. Genetic Algorithms, whereas other more expensive solutions such as e.g. minimax algorithms should be considered for our PSPACE-complete problems.

The rest of the paper is organized as follows. The following section introduces a motivation example informally introducing the key aspects of our setting. In Section~\ref{sec:proofs}, the model is formally introduced, and the complexity results are presented and proved. Conclusions and future work lines are given in Section~\ref{sec:conclusions}. The details of the proofs of our theorems are shown in the appendix of the paper.

Besides, although the aim of this paper is theoretical, as we are identifying the computational complexity of the problems under study, some considerations about how to deal with these problems in practice despite their high complexity are given in the appendix of the paper. A much more complex example than that given in the next section is outlined there and used to introduce these practical difficulties, as well as to discuss how to deal with them.

\section{Motivational example}
\label{sec:example}

In this section we use a simplified version of a robotic system to
introduce the main notions of our testing setting.
Let us test the {\it Collision Avoidance System} (CAS) of an autonomous four-wheeled exploration rover walking on potentially unstable and unbalanced surfaces. Three possible cases of interaction are considered in the CAS:

\begin{itemize}
    \item[A.-] If the collision object is expected to be reached in a time lower than one second, i.e.~{\it near} collision, then the robot fully turns to dodge it. This operation is assumed to have some risk, but it is considered the best choice because the robot is too close to fully brake before the collision.
    \item[B.-] If the collision object is expected to be reached in a time higher than two seconds, i.e.~{\it far} collision, then the robot brakes to fully stop before the collision. In this case, the alternative choice of turning is not taken because, compared to stopping, it is considered unnecessarily risky.
    \item[C.-] If the collision object is expected to be reached in a time between one and two seconds, i.e.~{\it intermediate} collision, then the robot is allowed to either fully brake or fully turn, as both choices are considered similarly risky.
    However, in this case the robot is not allowed to simultaneously brake {\it and} turn, since fully turning while the tires are blocked due to fully braking could make the robot overturn.
\end{itemize}

The tester fears the programming team may have made some mistakes while developing the robot, so the robot will be deployed in the laboratory and its behavior for different expected collision scenarios will be observed to rule out faults. In order to test the robot, the tester puts an object at some specific distance from the robot and makes the robot move towards it. Thus, the set of possible inputs produced by the tester to stimulate the robot (test cases) are those mentioned in cases (A)-(C), i.e., $I=\{near, far, inter\}$. Besides, the set of possible reactions of the robot (outputs) are defined as $O=\{nothing, turn, brake, both\}$. According to the limited freedom offered by previous conditions (A)-(C), as well as the mistakes the tester assumes the programmers may have made, it is constructed a set of possible   correct and incorrect full definitions of the IUT, i.e., definitions for all considered inputs. Each of these full definitions is formally described as a {\it function} where, for each possible input, the set of possible outputs that could be replied by the IUT after receiving that input is returned. When this set is a singleton, the behavior of that possible IUT definition for that particular input is deterministic else it is non-deterministic.

\begin{figure*}
    \hrule
    \scriptsize{
    $$\hspace*{-2em}\begin{array}{l}\begin{array}{llll}
    f_1(near)=\{turn\}&f_2(near)=\{turn\}&f_3(near)=\{turn\}&f_4(near)=\{turn\}\\
    f_1(far)=\{brake\}&f_2(far)=\{brake\}&f_3(far)=\{brake\}&f_4(far)=\{brake\}\\
    f_1(inter)=\{turn, brake\}&f_2(inter)=\{turn\}&f_3(inter)=\{brake\}&f_4(inter)=\{both\}\\
    \\
    f_5(near)=\{turn\}&f_6(near)=\{nothing\}&f_7(near)=\{turn\}&f_8(near)=\{brake\}\\
    f_5(far)=\{nothing\}&f_6(far)=\{brake\}&f_7(far)=\{turn\}&f_8(far)=\{brake\}\\
    f_5(inter)=\{turn\}&f_6(inter)=\{brake\}&f_7(inter)=\{turn\}&f_8(inter)=\{brake\}\\
    \\
    f_9(near)=\{nothing\}&f_{10}(near)=\{turn\}&f_{11}(near)=\{nothing\}\\
    f_9(far)=\{brake\}&f_{10}(far)=\{nothing\}&f_{11}(far)=\{nothing\}\\
    f_9(inter)=\{nothing\}&f_{10}(inter)=\{nothing\}&f_{11}(inter)=\{nothing\}\\
    \end{array}\\\\
    \begin{array}{llll}
    f_{12}(near)=\{turn\}&f_{13}(near)=\{nothing\}\\
    f_{12}(far)=\{nothing\}&f_{13}(far)=\{brake\}
    \\
    f_{12}(inter)=\{turn,nothing\}&f_{13}(inter)=\{nothing,brake\}
    \end{array}\end{array}$$
    }
    \hrule
    \caption{The possible definitions of the IUT. Definitions $f_1$-$f_3$ are considered correct, whereas $f_4$-$f_{13}$ are considered incorrect.}
    \label{fig:functions}
\end{figure*}

 \begin{figure*}[!t]
 \hrule

  \includegraphics[width=12cm]{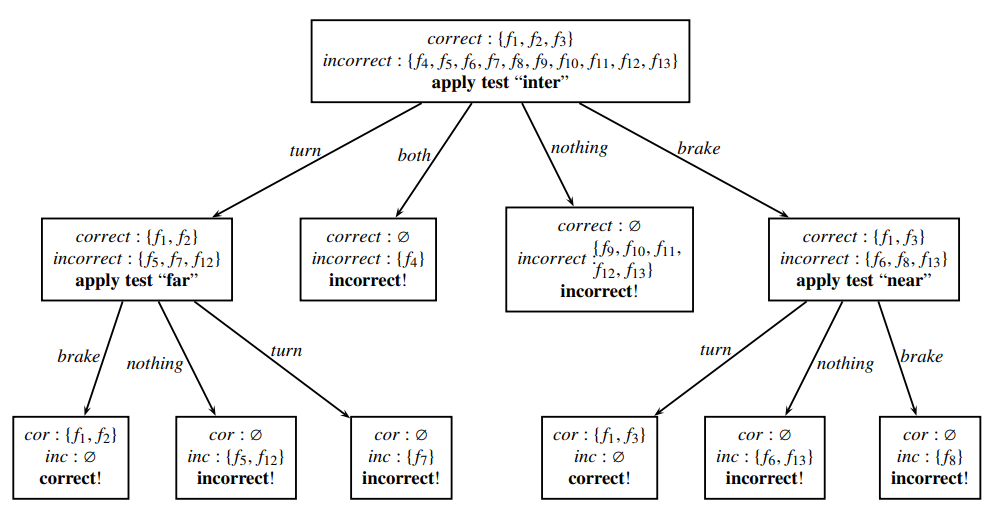}

   \hrule
  \caption{Adaptive testing strategy applying at most {\it two} test cases in all situations.}\label{fig:tree} 
 \end{figure*}

According to requirements (A)-(C), the possible IUT definitions $f_1$, $f_2$, $f_3$ depicted in Figure~\ref{fig:functions} are considered correct. Note that definition $f_1$ is allowed to non-deterministically brake or turn in intermediate-distance collisions (input $inter$), but this is allowed by requirement~(C). Also note that definitions $f_2$ and $f_3$ are fully deterministic. The tester fears the implementers may have made three possible kinds of mistakes:

\begin{itemize}
    \item[1.-] In intermediate-distance collisions, programmers could have made the robot {\it simultaneously} brake and turn.
    \item[2.-]
    Programmers may have incorrectly implemented the CAS signal requesting the robot to brake, making it do nothing in this case. Alternatively, this could have happened to the turning signal instead, or even simultaneously to both signals.
    \item[3.-] Programmers could have totally ignored the case distinction required by the CAS specification, making the robot brake in all situations. Alternatively, this could have happened with the turning operation instead of braking.
\end{itemize}

The possible incorrect IUT definitions resulting from these mistakes are shown in Figure~\ref{fig:functions}.
Mistake~(1) will be represented by the possible incorrect IUT definition~$f_4$. Besides, mistake~(2) is represented by several incorrect definitions resulting from altering correct definition~$f_1$ ($f_{12}$ and $f_{13}$), correct definition~$f_2$ ($f_5$ and $f_9$),  correct definition~$f_3$ ($f_6$ and $f_{10}$), and  any of them ($f_{11}$). Note that incorrect definitions $f_{12}$ and $f_{13}$ are non-deterministic. Finally, mistake~(3) is represented by incorrect functions $f_7$ and~$f_8$.

The previous tester assumptions about possible programmer mistakes are formalized by the tester hypothesis that the actual IUT definition belongs to set $\{f_1,\ldots,f_{13}\}$, consisting of those definitions fulfilling requirements (A)-(C) (i.e., $f_1,f_2,f_3$) and those failing them according to mistakes (1)-(3) (i.e., $f_4,\ldots,f_{13}$). The goal of the tester is interacting with the IUT, collecting observations, and using them to decide whether the IUT is necessarily correct (i.e., observations allow us to rule out all possible incorrect definitions) or incorrect (observations rule out all possible correct definitions). Assuming the tester can decide the next test cases to be applied depending on the observations collected by previous tests, i.e., adaptive testing, we tackle the following question: how many tests are needed in the worst case to decide the IUT \hbox{(in-)correctness?}

As shown in Figure~\ref{fig:tree}, we do not need to apply all three available test cases to make that decision, because just two of them are enough in any case if the second one cleverly depends on what is observed after the first one. First, we apply test case $inter$. If the output for this test is $both$, then the IUT is necessarily incorrect, as it must be $f_4$. If the output is $nothing$, then the IUT is necessarily incorrect too, as it must be some definition in set $\{f_9,\ldots,f_{13}\}$. However, if the output is either $turn$ or $brake$, then a second test case must be applied next. On the one hand, if it is $turn$ then test case $far$ is applied next, and all three possible subsequent replies of the IUT will let us decide whether the IUT is correct or not. On the other hand, if the output after test case $inter$ is $brake$, next we apply test case $near$, and again all three possible replies give enough information to decide whether the IUT is correct or not. Note that making such decisions does not require having enough information to uniquely identify which function defines the behavior of the IUT. For instance, in two cases depicted in Figure~\ref{fig:tree} in the first and fourth boxes at the bottom row, we provide a ``correct'' diagnosis and finish testing even though we do not know if the IUT is $f_1$ or $f_2$, or $f_1$ or $f_3$, respectively. Since both pairs of functions are correct and the set of possible incorrect definitions is empty in the respective boxes, distinguishing between both possible correct functions is  unnecessary to decide that the IUT must be correct. Analogously,  distinctions between {\it incorrect} definitions are also unnecessary when we have several incorrect definitions and no correct one. This is the case in the second and fifth boxes of the bottom row, as well as in the medium row's box reached after observing output $nothing$. Also note that a definition may simultaneously exist in multiple boxes after applying some test case if the definition is non-deterministic. For instance,  $f_1$, $f_{12}$, and $f_{13}$ can each be found in multiple boxes of the medium row.

It is worth noting that any alternative adaptive testing strategy starting with test case $near$ would need the application of {\it three} test cases in the worst case,
so the order in which test cases are applied in each branch of the tree shown
in Figure~\ref{fig:tree} does matter a lot.
After applying $near$, the IUT must be $f_6$, $f_9$, $f_{11}$, or $f_{13}$ if the reply is $nothing$ (i.e., incorrect IUT), it is $f_8$ if the reply is $brake$ (incorrect IUT), and otherwise (i.e., $turn$ is replied) it can be any of the remaining eight functions: $f_1$, $f_2$, $f_3$, $f_4$, $f_5$, $f_7$, $f_{10}$, $f_{12}$. Then, applying test case $far$ would not be enough to provide an \hbox{(in-)correctness} diagnosis yet: observing $nothing$ and $turn$ would certainly be, but obtaining $brake$ would not, as the remaining possible functions would be $f_1$, $f_2$, $f_3$, $f_4$, where we have three correct functions and an incorrect one. Thus, continuing our strategy with test case $far$ does not allow us to finish testing after two test cases. Let us see that replacing $far$ with $inter$ does not work either. Recall that, before the application of the second test, we had functions $f_1$, $f_2$, $f_3$, $f_4$, $f_5$, $f_7$, $f_{10}$, $f_{12}$. After applying $inter$, output $both$ would mean the IUT is $f_4$ (incorrect) and output $brake$ would imply the IUT is either $f_1$ or $f_3$ (correct), but output $turn$ would mean the IUT is $f_1$, $f_2$, $f_5$, $f_7$, or $f_{12}$, where we have two correct and three incorrect functions.

Symmetric arguments can be given to show that any testing strategy starting with test case $far$ will also need the application of three test cases in the worst case. Indeed, the strategy depicted in Figure~\ref{fig:tree} is the only one providing (in-)correctness diagnoses in just two test cases in the worst case, so it is optimal.

This strategy would still let us decide whether the IUT is correct or not after introducing some function redefinitions. For instance, let us redefine the behavior of $f_7$ for test case $far$ so that $f_7(far)=\{turn,nothing\}$, which makes this incorrect function non-deterministic. Then, the same adaptive testing strategy would let us decide whether the IUT is correct or not, and the diagram depicted in Figure~\ref{fig:tree} would apply to this alternative scenario as it is after adding $f_7$ to the set of possible incorrect definitions in the second box of the bottom row. Note that $f_7$ is {\it also} present in the incorrect set of the {\it third} box, and this remains unchanged. The incorrectness diagnosis given in that second box is still valid after the addition of $f_7$, as now we have zero correct functions and three incorrect ones. However, no adaptive or preset testing strategy would allow us decide the (in-)correctness of the IUT if $f_7$ was made non-deterministic in the following alternative way: $f_7(far)=\{turn, brake\}$. Note that, in this case, the incorrect function $f_7$ would not be distinguishable from correct functions $f_1$ and $f_2$, as $f_7$ could keep on replying $brake$ to test case $far$ no matter how many times it is applied to the IUT.

Restricted versions of our adaptive testing problem will also be studied in this paper, in particular by considering the following situations: assuming all functions must be deterministic, so that this would apply to our example after removing $f_1,f_{12},f_{13}$ and using the original definition of $f_7$; assuming that either the set of correct functions or the set of incorrect functions are singletons, so that this would be achieved e.g., by removing $f_2$ and $f_3$, or by removing all incorrect definitions but $f_7$; or by assuming both requirements together, e.g., by removing $f_1,f_3,f_{12},f_{13}$ and using the original definition of $f_7$. The complexity of the resulting four problems will be studied in the next section.

In order to introduce the difficulties of adaptive testing in bigger and more realistic scenarios, as well as to discuss how to deal with these difficulties in practice, a much much complex example is outlined in the appendix of the paper.

\section{Formal model and complexity results}\label{sec:proofs}

First, we introduce the notion of collection of definitions, which denotes
the set of all possible behavioral definitions the IUT could have,
each one
denoted by a function. We assume that $\calP(A)$ denotes the powerset of set $A$.

\bdfn\label{def:computation} {\bf (Collection of definitions)} Let $I$ be a finite set of inputs and $O$ be a finite
set of outputs. A \emph{collection of definitions} $C$ for
$I$ and $O$ is a finite enumeration of functions $f: I\longrightarrow \calP(O)$ where for all
$i\in I$ we have $f(i)\neq\emptyset$.

If $f(i)=\{o\}$, i.e., $f(i)$ is a singleton, sometimes we will just write $f(i) = o$. If $f(i)$ is a singleton for all $i\in I$, then we will say that $f$ is \emph{deterministic}. If all $f\in C$ are deterministic, then we say that $C$ is \emph{deterministic}.

We will assume that each function $f$ is extensionally defined. That is, for each input $i\in I$ we explicitly define the value $f(i)$.
\edfn

Each function in $C$ is the formal representation of some possible behavior of the IUT for all inputs: given a function $f\in C$, $f(i)$ represents the set of defined outputs we
can obtain after applying input $i\in I$ to the computation artifact
represented by $f$. Since $f(i)$ is a set, $f$ may represent a
non-deterministic behavior. For instance, in our motivation example, $f_1(inter)=\{\mathit{turn},\mathit{break}\}$,
so $f_1$ is non-deterministic, although $f_2$ is deterministic: $f_2(near)=\{\mathit{turn}\}$,
$f_2(far)=\{\mathit{break}\}$ and $f_2(\mathit{inter})=\{\mathit{turn}\}$.
Note that $C=\{f_1,f_2,f_3,\ldots,f_{13}\}$ in this example.
That is, we define the set $C$ extensionally by enumerating all its (finite) elements.

Collections of definitions are used to represent the set of
possible implementations we are considering in a given testing scenario.
Implicitly, a collection of definitions $C$ represents the \emph{hypotheses}
about the IUT the tester is assuming, see e.g.,~\cite{hie02b}. For instance, if the tester assumes the IUT is deterministic then $C$ should only include deterministic functions; if the IUT assumes the IUT may produce wrong outputs for at most one input, then all functions in $C$ should be defined this way; etc.

Collections of definitions are also used to represent the subset of
specification-compliant implementations. Let $C$ represent the set of
possible implementations and $E\subseteq C$ represent the set of
implementations fulfilling the specification. The goal of testing is
interacting with the IUT so that, according to the collected
responses, we can decide whether the IUT actually belongs to $E$ or
not. Typically, we apply some tests, i.e., some inputs
$i_1,i_2,\ldots\in I$, to the IUT one after each other
so that the observed results $o_1\in f(i_1)$, $o_2\in f(i_2)$, $\ldots$
allow us to provide a verdict.

\bdfn\label{def:specification} {\bf (Specification and testing scenario)} Let $C$ be a collection of definitions. A
\emph{specification} of $C$ is a set $E\!\subseteq\! C$.
A \emph{testing scenario} is a tuple $(C,E)$.
\edfn 

If $f\in E$ then $f$ denotes a \emph{correct} behavior, while $f\in
C\backslash E$ denotes that $f$ is incorrect. Thus, a specification $E$
implicitly denotes a correctness criterion, and the set $C\backslash E$ implicitly defines the \emph{fault model} assumed by the tester, as it consists of all incorrect behaviors the IUT may have. In any case, the correctness of a function in our model consists, by definition, in belonging to $E$,
where $E$ is extensionally defined function by function.
In our motivation example, $E=\{f_1,f_2,f_3\}$, and this set reflects the specification implicitly defined by the requirements (A)-(C) mentioned in Section~\ref{sec:example}. Besides, $C\backslash E = \{f_4,\ldots,f_{13}\}$, and this set reflects the fault model implicitly given by the set of considered mistakes (1)-(3) in that section. Alternatively, let a specification require that a program, receiving and producing natural numbers, computes the square of the input value, and let our fault model assume that each possible wrong program $P_i$, with $i\in [-10,10]$, will produce outputs being $i$ units above the actual square. Then, $E=\{f\}$ with $f(x) = x^2$, and $C = \{g_i\;|\; i\in[-10,10]\}$ where $g_i(x) = f(x) + i$.

When we deal with adaptive testing, we will not use a set $\calI\subseteq I$ to denote a possible test suite, because now we are free to select our second input test after observing the output obtained after applying the first input, to select our third input test after obtaining the second output, and so on. Thus, we will rather speak about testing {\it strategies}, which are neither sets nor sequences. A testing strategy can be seen as a {\it tree} where each node represents the application of a test, the number of children of each node is the number of possible outputs that can be obtained for such input, and verdicts of necessary (in-)correctness are given in the leaves of the tree, as depicted in Figure~\ref{fig:tree}.
Thus, the sequence of nodes of each branch from the root to a leaf of the tree represents a possible sequence of tests to be applied to a IUT (similarly to a test suite in preset testing), but the concrete branch to be applied to each IUT depends on the sequence of outputs produced by the IUT during the testing process.

The aim of our problem is to decide whether it is possible or not to find a testing tree such that its depth, i.e.,~the number of input tests of its largest branch, is at most $k$, being $k$ a given natural number.
Let us start defining the decision problem in case we are only dealing with deterministic functions, that is, for all $f \in C$ and $i\in I$, $f(i)$ is a singleton.

\bdfn\label{def:DD} {\bf (Deterministic Adaptive Test Decision Problem)} Let $C$ be a deterministic collection of definitions for $I$ and $O$, $E\subseteq C$ be a specification, and $k$ be a natural number. We say that $(C,E,k)$ satisfies $\DD$ if:

\hspace*{-1em}$
\begin{array}{c}
\exists i_{1}\in I\,\,\forall o_{1}\in\{f(i_{1}):f\in C\}\\
\exists i_{2}\in I\,\,\forall o_{2}\in\{f(i_{2}):f\in C,\,\,o_{1}=f(i_{1})\}\\
\vdots\\
\exists i_{l}\in I\,\,\forall o_{l}\in\{f(i_{l}):f\in C,\,\,o_{m}=f(i_{m})\,\,\forall m<l\}\\
\vdots\\
\exists i_{k}\in I\,\,\forall o_{k}\in\{f(i_{k}):f\in C,\,\,o_{m}=f(i_{m})\,\,\forall m<k\}\\
\left\{ f\in C:o_{m}=f(i_{m})\,\,\,\forall m\leq k\right\} \subseteq E\,\,\vee\\
\left\{ f\in C:o_{m}=f(i_{m})\,\,\,\forall m\leq k\right\} \subseteq C\backslash E\\[3mm]
\end{array}$

Given $(C,E,k)$, the {\it Deterministic Adaptive Test Decision Problem} consists in finding out whether $(C,E,k)$ satisfies $\DD$. In a notation abuse, the problem itself will also be denoted by $\DD$. In $\DD$ instances, we will assume $k$ is given in binary form.

When either $E$ is a singleton or $C\backslash E$ is a singleton, that is, there is only one correct or there is only one incorrect function, the resulting particular-case problem will be denoted by $\DD_1$.
\edfn

In the previous definition, satisfying $\DD$ actually consists in finding an adaptive testing strategy applying no more than $k$ inputs.\footnote{Although the definition requires exactly $k$ inputs, using less inputs is clearly allowed, as filler inputs could be trivially added to satisfy the definition, e.g., repeated inputs.}
Each new level of possible input tests depends on the previous outputs produced by the IUT. Moreover, for each branch of inputs of the strategy from the root to a leaf, the functions $f\in C$ that are consistent with the outputs observed through the branch (that is, $\{ f\in C:o_{m}=f(i_{m})\,\,\,\forall m\leq k\}$) either belong all to the set of correct functions $E$ or belong all to the set of incorrect functions $C\backslash E$. Thus, each branch of inputs classifies the corresponding IUT as either correct or incorrect.
In addition, as the branches consider all possible outputs produced by the IUT, the whole tree of test cases is actually {\it complete} in the task of determining whether the IUT is correct or not.

Once the decision problem has been defined, it is trivial to define the corresponding optimization problem, where we are interested in finding the minimum $k$ such that $(C,E,k)$ satisfies $\DD$.

\bdfn\label{def:DO} {\bf (Deterministic Adaptive Test Optimization Problem)} Let $C$ be a deterministic collection of definitions for $I$
and $O$, and $E\subseteq C$ be a specification. Given $(C,E)$, the {\it Deterministic Adaptive Test Optimization Problem} ($\DO$) consists in finding
$\min\{k:\,(C,E,k)$ satisfies $\DD\}$.

When $E$ is a singleton or $C\backslash E$ is a singleton, we will use $\DO_1$ to refer to $\DO$.
\edfn

Once we have defined the decision and optimization problems in the deterministic scenario, we can easily modify them to deal with the non-deterministic case. In order to do that, we only need to take into account that the output obtained by each function is not unique. Thus, we have to consider the union of all possible outputs produced by all possible functions at each of the levels of the corresponding tree, as shown in the next definition.

\bdfn\label{def:ND} {\bf (Non-deterministic Adaptive Test Decision Problem)} Let $C$ be a collection of definitions for $I$
and $O$, $E\subseteq C$ be a specification, and $k$ be a natural number. We say that $(C,E,k)$ satisfies $\ND$
if:

$
\begin{array}{c}
\exists i_{1}\in I\,\,\,\forall o_{1}\in\underset{f\in C}{\cup}f(i_{1})\\
\exists i_{2}\in I\,\,\,\forall o_{2}\in\underset{f\in\left\{ f\in C:\,o_{1}\in f(i_{1})\right\} }{\cup}f(i_{2})\\
\vdots\\
\exists i_{l}\in I\,\,\,\forall o_{l}\in\underset{f\in\left\{ f\in C:\,o_{m}\in f(i_{m})\,\forall m<l\right\} }{\cup}f(i_{l})\\
\vdots\\
\exists i_{k}\in I\,\,\,\forall o_{k}\in\underset{f\in\left\{ f\in C:\,o_{m}\in f(i_{m})\,\forall m<k\right\} }{\cup}f(i_{k})\\
\left\{ f\in C:o_{m}\in f(i_{m})\,\,\,\forall m\leq k\right\} \subseteq E\,\,\vee\\
\left\{ f\in C:o_{m}\in f(i_{m})\,\,\,\forall m\leq k\right\} \subseteq C\backslash E \\[3mm]
\end{array}$

Given $(C,E,k)$, the {\it Non-deterministic Adaptive Test Decision Problem} consists in finding out whether $(C,E,k)$ satisfies $\ND$. The problem itself will also be denoted by $\ND$. In $\ND$ instances,  $k$ will be defined in binary form.

When either $E$ or $C\backslash E$ is a singleton, the resulting problem will be denoted by $\ND_1$.
\edfn

Again, it is trivial to define the corresponding optimization problem, as shown in the next definition.

\bdfn\label{def:NO} {\bf (Non-Deterministic Adaptive Test Optimization Problem)} Let $C$ be a collection of definitions for $I$
and $O$, and $E\subseteq C$ be a specification. Given $(C,E)$, the {\it Non-deterministic Adaptive Test Optimization Problem} ($\NO$) consists in finding
$\min\{k:\,(C,E,k)$ satisfies $\ND\}$.

When $E$ is a singleton or $C\backslash E$ is a singleton, we will use $\NO_1$ to refer to $\NO$.
\edfn

Given the obvious equal treatment of $E$ and $C\backslash E$ in the definitions of the satisfaction conditions of $\DD$ and $\ND$ (see definitions~\ref{def:DD} and~\ref{def:ND}), hereafter only the case where $E$ is a singleton will be regarded when dealing with all problems $\DD_1$, $\DO_1$, $\ND_1$, and $\NO_1$.

Once we have defined our problems, we present the following properties for the deterministic scenario. The inclusion of $\DD$ in class $\NP$ might look strange at a first glance, given the apparent exponential size of $\DD$ solutions, as they are trees, and, moreover, the apparent exponential {\it depth} of these trees with respect to the size representation of $\DD$ instances, as $k$ is defined in {\it binary} in them.

\bthm \label{th:Det} We have:
\begin{itemize}
    \item[(a)] $\DD$ is $\NPc$.
    \item[(b)] $\DD_1$ is $\NPc$.
    \item[(c)] $\DO$ is $\LogAPXh$.
    \item[(d)] $\DO_1$ is $\LogAPXc$.
\end{itemize}

\ethm

\bprf
Let us start proving (c).
In order to do it, we have to provide a Log-APX-hardness preserving polynomial reduction from a $\LogAPXh$ problem into $\DO$. We consider an S-reduction\footnote{In an S-reduction, the solutions of the second problem can be mapped into solutions of the first problem having exactly the same cost, and the optima of both problems have the same cost as well. See e.g.,~\cite{Crescenzi97}.} from {\it Minimum Set Cover}, $\MSC$~\cite{feige1998threshold}.
Given a collection $\calC$ of
sets $\calC=\{S_1,\ldots,S_m\}$ with $\bigcup_{S\in \calC} S =
\{e_1,\ldots,e_p\}$, $\MSC$ consists in picking a subset $\calC'$ of
$\calC$ such that $\bigcup_{S\in \calC'} S = \{e_1,\ldots,e_p\}$ in
such a way that $|\calC'|$ (i.e., the number of sets in $\calC'$) is
minimized. The construction of an S-reduction from $\MSC$
into $\DO$ can be done as follows. From an $\MSC$ instance $\calC$ defined as before, we
define a $\DO$ instance in the same terms as in
Definition~\ref{def:DO} where we will have a different input for each set, a single correct function (which always returns output $0$), and an incorrect function for each possible element of the subsets:
\begin{itemize}
\item $I = \{1, 2, \ldots, m\}$
\item $O = \{0,1\}$
\item $E=\{g\}$, where $\forall i\in I ~ g(i)=0$
\item $C=E\cup \{f_e : e \in \{e_1,\ldots,e_p\} \}$
where
$$f_{e}(l) = \left\{\begin{array}{ll}1 & \si e \in S_l\\0 &  \eoc  \end{array}\right.$$
\end{itemize}

Let us remark that each input test represents a set of the original $\MSC$ problem. Moreover, there is an incorrect function for each possible element of the original sets, and the application of this function to an input (that is, to a \emph{set}) returns $1$ iff the element belongs to the corresponding set.

We can see that there exists a solution to this $\DO$ instance with
cost $q\in\nat$ (say solution $\{S_{l_1},\ldots,S_{l_q}\}$) iff there exists a solution to the original $\MSC$
instance with the same cost $q$, and we can easily map the former into the latter.

Taking into account our reduction, $\bigcup_{j=1}^q S_{l_j}$ is a set cover of $\calC$ iff $\forall e\in \bigcup \calC ~ \exists r\in\{1\ldots q\}$ such that $e\in S_{l_r}$. Thus, $f_e(l_r)=1$. Hence, by using input $l_r$ in $\DO$ we {\it distinguish} function $f_e$ from the only correct function $g$, which always returns $0$, i.e., the observed output will always let us discard one of them. Therefore, the set of inputs $\{l_j : j\in\{1,\ldots,q\}\}$ allows to distinguish $g$ from all functions in $C\backslash \{g\}$. Hence, by applying these $q$ inputs in any order, observed outputs will always let us know either the IUT is correct or incorrect.

Analogously, if some set of $q$ inputs $\{{l_j}:j\in\{1,\ldots,q\}\}$ allows to distinguish $g$ from all incorrect functions, then
for each incorrect function $f_e$ we have an input $l_r$ such
that $f_e(l_r)=1$, because the correct function always returns $0$.
Thus, in the corresponding $\MSC$ problem we will also have that
for each element $e\in\bigcup\calC$ we have a set $S_{l_r}$ such
that $e\in S_{l_r}$. Thus,
$\{S_{l_j}:j\in\{1,\ldots,q\}\}$ is a set cover of $\calC$ with cost $q$.

We conclude that there is an S-reduction from $\MSC$ into $\DO$, and this reduction
proves that $\DO$ is $\LogAPXh$.
Moreover, note that the mapping proposed in our S-reduction also polynomially reduces $\MSC$ to $\DD$, which proves that $\DD$ is $\NPh$. In order to conclude (a), we also need to show $\DD$ is $\NP$.

Let us recall our view of testing strategies as trees, as illustrated in Figure~\ref{fig:tree}.
Since all functions are deterministic in $\DD$ (contrarily to the case shown in Figure~\ref{fig:tree}, actually dealing with $\ND$), the sets of all (correct and incorrect) functions being possible at the children nodes of a given node of a testing strategy must be pairwise disjoint. By a trivial inductive reasoning, this implies that the sets of functions being possible at all leaves of the strategy must be pairwise disjoint as well, and in fact they must constitute a partition of the collection of definitions~$C$. Let $n = |C|$ denote the number of functions in $C$ and $m = |I|$ denote the number of inputs.
We infer that the number of leaves of a testing strategy for $\DD$ cannot be higher than $n$.
Let us consider a testing strategy where all applied inputs are different from all other inputs appearing in the {\it same} branch, i.e., no branch of the strategy contains repeated inputs. Clearly, the depth of this strategy cannot be higher than $m$. Note that deciding $\DD$ does not require exploring any testing strategy where some branch contains repeated inputs, because repeated inputs are useless. This is obvious given the deterministic nature of all functions in $\DD$. Hence, regardless of the value of $k$ given in a $\DD$ instance $(C,E,k)$, we can always decide $\DD$ by considering  trees with no more than $n$ leaves and no more than $m$
depth. We infer that the total number of nodes of any candidate tree we might need to consider to decide $\DD$  cannot be higher than $n\cdot m$.
Moreover, as the representation size of $C$ in a $\DD$ instance cannot be lower than $n$ and cannot be lower than $m$, we infer that the sizes of candidate solutions of $\DD$, i.e., testing strategies, are polynomial with the sizes of the corresponding problem instances. Given this limitation, it is easy to see that checking the validity of any candidate solution takes polynomial time, so $\DD$ is $\NP$.

Let us note that, in our previous S-reduction, we produced $\DO$ instances which are also, in fact,  $\DO_1$ instances, and the cost functions to be minimized in both optimization problems coincide. Thus, that S-reduction also proves that $\DO_1$ is $\LogAPXh$ and $\DD_1$  is $\NPh$.

As we have 
proven $\DO_1$ is $\LogAPXh$, in order to prove (d) we only need to prove $\DO_1$ is $\LogAPX$. We can do it by providing an S-reduction from $\DO_1$ into $\MSC$. As we know $\MSC$  is $\LogAPX$, this will imply that $\DO_1$ is also in $\LogAPX$.

The construction of an S-reduction from $\DO_1$
into $\MSC$ can be done as follows. From a $\DO_1$ instance ($C,\{g\})$, we
define an $\MSC$ instance  where we create a set $S_i$ for each input $i\in I$. The elements of the sets will be the incorrect functions of the original $\DO_1$ problem. Moreover, each set $S_i$ will contain those functions that produce an incorrect output for input~$i$. That is:
$$S_i = \{f : f \in C\backslash \{g\}, f(i) \neq g(i)\} ~~\forall i\in I$$

Then, a set $\{S_{l_i}:i\in\{1,\ldots,q\}\}$ is a set cover iff each function $f\in C \backslash \{g\}$ is included in at least a set $S_{l_i}$. This property holds iff for each function $f\in C \backslash \{g\}$ there exists at least one input $l_i$ such that $f(l_i)\neq g(l_i)$, meaning that applying all inputs in the set
$\{l_i:i\in\{1,\ldots,q\}\}$ in any order is a valid testing strategy for the original testing problem.

Let us remark that the cost of the solution is the same in both problems, $q$, also when the solution of $\MSC$ is optimal. So, we have an  S-reduction from $\DO_1$ to $\MSC$. Thus, $\DO_1$ is     $\LogAPX$. As we have already proved  $\DO_1$ is $\LogAPXh$, we conclude that $\DO_1$ is $\LogAPXc$.

Moreover, by taking this S-reduction as a polynomial reduction, we also prove $\DD_1$ is $\NP$, as $\MSC$ is $\NP$. Since we proved earlier that $\DD_1$ is $\NPh$, we conclude $\DD_1$ is $\NPc$, as stated in~(b).
\eprf
\vspace*{1em}

Once we have analyzed the deterministic scenario, next we deal with the non-deterministic case.

\bthm \label{th:NonDet}
We have that
$\ND$ and $\ND_1$ are $\PSPACEc$.
\ethm

\bprfsketch%
%
The complete proof 
can be seen in the appendix, here we describe the main ideas. First, we prove that $\ND$ is $\PSPACE$
and, as a consequence, we will also have that its particular case $\ND_1$ is also in $\PSPACE$. Then, we prove that $\ND_1$ is $\PSPACEh$ and, as a consequence, we will also have that its generalization $\ND$ is also $\PSPACEh$.

{\bf $\ND$ is $\PSPACE$}

Let us show an algorithm that solves $\ND$ by using a polynomial amount of
memory with respect to the size of the input.
We assume that there is a total order for the elements of $I$ and
also a total order for the elements of $O$, and that we can compare any pair of elements using a polynomial amount of memory with respect to the size of the sets $I$ and $O$. This can be trivially done, as we can explicitly define all the possible pairs by using $|I|^{2}$ and $|O|^{2}$ memory.

Our algorithm will use two vectors $i$ and $o$, as well as a natural variable $l$. Vector $i$ (respectively, $o$) will contain the input (resp. output) variables we are currently considering at the candidate testing strategy branch under study, while the natural variable $l$ will contain the depth we are located at in the decision tree. Assuming the testing strategy is allowed to apply up to $k$ inputs in our $\ND$ instance, apparently $k$ should be the size of vectors $i$ and $o$, as well as the maximum possible value of $l$. However, note that the amount represented by $k$ is exponential with respect to the length of the binary code representing it in the $\ND$ instance. Thus, the size of vectors $i$ and $o$ need to be smaller than $k$ to guarantee a polynomial use of memory. Their actual size will be explained later.

The algorithm, whose pseudocode is given in Appendix~\ref{ap:algoritmo}, tries all  possible combinations of inputs and outputs in a minimax fashion. %
In case a combination does not allow to
either discard all correct functions or discard all incorrect functions, then this combination is not useful. Thus, we modify the last input whose possible values have not been completely tested yet, that is, we try the next input possibility. In case there are not more input possibilities, it means that the answer to the decision problem is not.

In case a given combination allows to discard all correct functions or all incorrect functions, then we change the last output whose possible values have not been completely tested, that is, we go on to check whether the new output does not modify the feasibility of this strategy.
In case there are not more output possibilities, it means that we have been able to discard all correct or all incorrect functions in all the possible situations. That is, the answer to the decision problem is yes.

Despite the exponential size of the tree being explored, during the execution of the algorithm the program only needs to remember where it is along the tree exploration, and vectors $i$ and $o$ and natural number $l$ define that place.
Note that, regardless of the actual value of $k$ given in the $\ND$ instance $(C,E,k)$, no testing strategy constructed for deciding $\ND$ needs to have a depth higher than $h$, where $h = |I|$ is the number of available inputs, because repeating inputs in the same branch is useless. Notice that the non-determinism of functions in $\ND$ allows the IUT to reply the same output to the same input no matter how many times it is applied. Hence, if some node of the testing strategy applies a repeated input, i.e., an input already applied in some previous node of the same branch, then, for some child of this node, the set of non-discarded functions will be exactly the same as in its parent node. Therefore, the problem of discarding either all correct or all incorrect functions available at the former node will be the same as that problem for the latter node ---although after applying one additional input, which makes this repeated input useless. We conclude that the depth of the exploration performed by the algorithm will never need to be higher than $h$, so the actual size of vectors $i$ and $o$, as well as the maximum possible value of $l$, can be defined as $min(k,h)$.
Since the number of available inputs $h$ cannot be higher than the representation size of $C$ in any $\ND$ instance $(C,E,k)$, the depth of the tree being explored is polynomial, and vectors $i$ and $o$ can be represented with polynomial memory.

In addition, each specific action of the algorithm requires a constant number of auxiliary variables. Note that we only need to obtain the minimum or maximum of a given set, to obtain the next element of a set, to check whether $o[m]\in f(i[m])$
for all $1\leq m\leq min(k,h)$, and to check whether $f \in E$ or $f \in C\backslash E$.
We conclude that the algorithm runs with polynomial space.

{\bf $\ND_1$ is $\PSPACEh$}

In this case we have to provide a polynomial reduction from a $\PSPACEh$ problem into our problem $\ND_1$. In particular, we will reduce the {\it True Qualified Boolean Formula} problem ($\TQBF$~\cite{sipser12,stockmeyer87}) into $\ND_1$.
In this problem, given a propositional logic formula and a quantification of all its variables by universal and existential quantifiers, we decide if the resulting first-order logic formula is true, i.e., equivalent to $\top$.
There are several possible presentations of this problem, and in particular we assume that the propositional formula is given in {\it conjunctive normal form} (CNF)
and that existential and universal quantifiers alternate. That is, formulas will always have the following form:

$$
\exists x_{1}\forall y_{1}\exists x_{2}\forall y_{2}\ldots\exists x_{k}\forall y_{k}\,\,\phi
$$

\noindent where $\phi$ is a propositional logic formula  defined over variables $x_1,y_1,\ldots,x_k,y_k$ which is given in CNF, i.e., $\phi = c_1 \y \ldots \y c_n$ for some disjunctive clauses $c_1,\ldots,c_n$.

Our polynomial reduction from $\TQBF$
into $\ND_1$ will be easier to introduce if we view both problems as games: in $\TQBF$ a ``player'' tries to iteratively set the existential variables in such a way that the formula will be true no matter how the corresponding universal variables are set by an ``opponent,'' and in $\ND_1$ the tester iteratively selects the inputs to be applied to the IUT in such a way that either all correct or all incorrect functions are eventually discarded no matter how the corresponding outputs are  chosen by the  ``opponent IUT.'' Our reduction will map $\TQBF$ existential variables into $\ND_1$ inputs (chosen by the tester), and $\TQBF$ universal variables into the corresponding $\ND_1$ outputs produced afterwards (chosen by the IUT). Hence, when the tester introduces some input implicitly setting the value of existential variable $x_j$, the IUT will reply the output implicitly setting the value of  universal variable $y_j$. As we will see, formula $\phi$ will be true under some variable values iff either all correct or all incorrect functions are discarded under the corresponding inputs and outputs. In particular, each disjunctive clause $c_i$ of $\phi$ in $\TQBF$ will be represented by an {\it incorrect} function $f_{c_i}$ in $\ND_1$ (actually, satisfying $c_i$ will be equivalent to discarding $f_{c_i}$), and the $\ND_1$ instance will contain a single correct function $g$. Since the tester will immediately {\it win} if the only correct function is discarded, the hard and most interesting part of classifying the IUT as correct or incorrect {\it in all cases} will be when IUT outputs are always consistent with that single correct function.

More technically, from a $\TQBF$ instance of the form $\exists x_1\forall y_1 \ldots\exists x_k\forall y_k \phi$, we
define a $\ND_1$ instance  where we will have: two different inputs for each variable existentially quantified, one representing that $x_j$ is set to $\top$ and another one representing it is set to $\bot$; a single correct function $g$, which can return $0$ or $1$ for all inputs; three possible outputs $0,1,-1$ representing the two possible values of variables $y_j$ and a new extra value ($-1$) to distinguish any function producing it from the correct function, which cannot produce it; and an incorrect function ($f_{c_i}$) for each disjunctive clause ($c_i$) of the logic formula $\phi$. Formally,
\begin{itemize}
\item The allowed depth of the testing strategy, i.e., third value of our $\ND_1$ instance, is set to $k$.
\item $I = \{x_1, x_2, \ldots, x_k\} \cup \{\overline{x_1}, \overline{x_2}, \ldots, \overline{x_k}\}  $,
\item $O = \{0,1,-1\}$,
\item $E=\{g\}$, where $\forall i\in I ~ g(i)=\{0,1\}$,
\item $C=E\cup \{f_{c_i}: c_i \in \phi \}$
where
$$f_{c_i}(x_j) = \left\{\begin{array}{ll}
-1 & \si x_j \in  c_i \\
-1 & \si y_j, \overline{y_j} \in c_i\\
0 & \si y_j \in c_i \y x_j,\overline{y_j}\notin c_i\\
1 & \si \overline{y_j} \in c_i \y x_j, y_j\notin c_i\\
\{0,1\} &  \eoc
\end{array}\right.
$$

$$f_{c_i}(\overline{x_j}) = \left\{\begin{array}{ll}
-1 & \si \overline{x_j} \in  c_i \\
-1 & \si y_j, \overline{y_j} \in c_i\\
0 & \si y_j \in c_j \y \overline{x_j},\overline{y_j}\notin c_i\\
1 & \si \overline{y_j} \in c_i \y \overline{x_j}, y_j\notin c_i\\
\{0,1\} &  \eoc
\end{array}\right.
$$

\end{itemize}

Each function $f_{c_i}$ will be discarded with some inputs and outputs in the $\ND_1$ instance
iff clause $c_i$ is satisfied in the original $\TQBF$ instance with the corresponding variable values. If  literal $x_j$ (analogous for $\overline{x_j}$) appears in disjunctive clause $c_i$, then $c_i$ is obviously satisfied when $x_j$ holds. Then, input $x_j$ (respectively, $\overline{x_j}$) distinguishes $f_{c_i}$ from the correct function $g$, because $f_{c_i}$ returns~$-1$ to that input and $g$ returns either $0$ or $1$. Note that if the IUT returns $-1$ then the tester immediately  {\it wins} by discarding the only correct function $g$, and if it returns $0$ or $1$ then the tester will discard $f_{c_i}$. In addition, if both $y_j$ and $\overline{y_j}$ belong to $c_i$ then $f_{c_i}$ will return $-1$ for both inputs $x_j$ and $\overline{x_j}$, as $c_i$ will be trivially satisfied in both cases. Otherwise, if literal $y_j$ (analogous for $\overline{y_j}$) appears in clause $c_i$,
then $f_{c_i}$ is defined to return the {\it opposite} value to that fulfilling that literal. For instance, if literal $\overline{y_j}$ appears in clause $c_i$ and the IUT replies $0$ to $x_j$ or $\overline{x_j}$, i.e., the IUT is setting $y_j$ to $\bot$, then $f_{c_i}$ is {\it discarded}, i.e., clause $c_i$ is satisfied, because $f_{c_i}$ is defined to return $1$ in that case.

The previous construction is not correct yet as it, because we must still solve two problems. On the one hand, although only $k$ inputs are allowed in the constructed $\ND_1$ instance, the tester could apply $k$ inputs and still represent an impossible logical valuation. For instance, for $k=3$, the tester could apply $x_1$, $\overline{x_1}$, and $x_3$, thus giving both possible values to $x_1$ and leaving $x_2$ undefined. On the other hand, we must guarantee that the order used to select the inputs (i.e., tests) is the same as the order in which variables appear in the corresponding $\TQBF$ formula. Note that inputs (i.e., tests) can be applied in any order in $\ND_1$, but variables are introduced in a mandatory order in $\TQBF$, and this order does matter. For instance, it is easy to see that $\exists x_1 \forall y_1 \exists x_2 \forall y_2 ((x_2 \y y_1) \o (\overline{x_2} \y \overline{y_1}))$ holds but $\exists x_2 \forall y_2 \exists x_1 \forall y_1 ((x_2 \y y_1) \o (\overline{x_2} \y \overline{y_1}))$ does not.

We will add new incorrect functions to be discarded to our $\ND_1$ instance, and we will make sure that it is impossible to discard all of them {\it in all cases} (more precisely, in all cases where the tester does not trivially win by discarding the only correct function $g$) {\it unless} exactly one valuation is given by the tester to each existential variable (the first problem mentioned above) and all of them are applied in the order required by the $\TQBF$ instance (the second problem). An additional incorrect function $f_0$, as well as new incorrect functions $f_j$ and $f'_j$ for all $1 \leq j \leq k-1$, will be added to set $C$. By their design, discarding all of them will force the tester to apply at least one input representing each existential variable, it does not matter if positively or negatively. Given the limit of $k$ inputs to be applied in our $\ND_1$ instance, this will solve the first of our previous problems.

In order to solve the second problem, i.e.~the order, the new incorrect functions will be defined as follows. For any input representing a value for existential variable $x_j$ (again, it does not matter if positive or negative), $f_j$ and $f'_j$ will  reply $0$ and $1$, respectively.
Given this definition, the IUT will discard $f_j$ if it replies $1$ and $f'_j$ if it replies $0$. Alternatively, in any branch where the IUT replies $-1$, $g$ will be discarded and the tester will trivially win, so next we focus on all branches where this does not happen. Note that the choice between discarding $f_j$ or discarding $f'_j$ will be out of the control of the tester, as this will be decided by the IUT. We will give the tester the power to control which one of either $f_j$ or $f'_j$ is discarded when the input corresponding to variable $x_{j+1}$ is applied. In order to allow this, we duplicate all the inputs representing existential variables into {\it prime} and {\it non-prime} versions, so that for each $j$ we will have versions $x_j$, $\overline{x_j}$, $x'_j$, and $\overline{x_j}'$. When a version of the input corresponding to $x_{j+1}$ is applied to the IUT, function $f'_j$ will be necessarily discarded if a non-prime version is used, i.e., $x_{j+1}$ or $\overline{x_{j+1}}$, as $f'_j$ will return $-1$ to any of them, and function $f_j$ will be discarded if a prime version is used (i.e., $x'_{j+1}$ or $\overline{x_{j+1}}'$), for which $f_j$ will return $-1$. The choice of the version of $x_{j+1}$ to be applied will let the tester control which one of $f_j$ or $f'_j$ is discarded but, again, any version of $x_{j+1}$ will discard either $f_{j+1}$ or $f'_{j+1}$ {\it without} any tester control. Therefore, the tester will be forced to discard the non-discarded function with a suitable version of $x_{j+2}$, and so on.

We can see that discarding {\it all} functions $f_0$, $f_1$, $f'_1$, $\ldots$, $f_{k-1}$, $f'_{k-1}$ will force the tester to apply the chosen versions of variables $x_1,\ldots,x_k$ in that {\it exact} order, while $f_0$, the anchor function of this process, will be discarded by any version of $x_1$. If some version of variable $x_{j+1}$ is applied to the IUT {\it before} the corresponding version of $x_j$ is applied, then when we apply the former we will not know which one of either $f_j$ or $f'_j$ will not be discarded {\it afterwards} by the latter, so the decision of which function needs to be {\it rescued} by $x_{j+1}$ will be blind. Hence, there will exist a branch in the testing strategy where the application of $x_j$ discards, between $f_j$ and $f'_j$, the function which was {\it already} discarded before, letting the other one non-discarded and making the tester lose.

By selecting some input $x_j$, $\overline{x_j}$, $x'_j$, or $\overline{x_j}'$ for each $j$, the tester will simultaneously make a decision in the negation vs non-negation axis to discard all $f_{c_i}$ functions, and in the prime vs non-prime axis to discard all $f_j$ and $f'_j$ functions. Note that the tester will {\it not} win if there exists a branch in the testing strategy where $g$ and either some $f_{c_i}$ or some $f_l$ or $f'_l$ is not discarded. Thus, satisfying all clauses {\it and} keeping the variables order with a consistent valuation of variables are both mandatory conditions to win in the $\ND_1$ instance.

Given the previous considerations, the $\ND_1$ instance its now defined as follows. The correctness of the resulting polynomial reduction is proved in Appendix B. 

\begin{itemize}
\item $I = \{x_1, x_2, \ldots, x_k\} \cup \{\overline{x_1}, \overline{x_2}, \ldots, \overline{x_k}\}  \\
\cup \{x'_1, x'_2,\ldots, x'_k\} \cup \{\overline{x_1}', \overline{x_2}', \ldots, \overline{x_k}'\}
$,
\item $k$, $O$, and $E$ are not modified. Again, $g(i)=\{0,1\}$ for all inputs $i\in I$, including the new ones,
\item $C=E\cup \{f_{c_i}: c_i \in \phi \}
\cup \{f_0\} \cup \{f_l, f'_l : \forall l \in \{1,\ldots,k-1\}\}$
where $f_{c_i}$ is defined as before for all $x_j$ or $\overline{x_j}$, and the values it returns for the new inputs $x'_j$ (analogously $\overline{x_j}'$) are the same as those returned by $x_j$ (analogously, $\overline{x_j})$, that is,\\

$f_{c_i}(x'_j) = f_{c_i}(x_j)$ ~~~~
$f_{c_i}(\overline{x_j}') = f_{c_i}(\overline{x_j})$\\


Regarding the new functions $f_l$, we use $2k-1$ of them: two for each existentially quantified variable $x_j$ ($f_j$ and $f'_j$) but the first one, for which we add only one function ($f_0$).
These functions are defined in the same way for each variable and its negation, that is:\\

$\begin{array}{ll}
f_l(x_j) = f_l(\overline{x_j}) & f_l(x'_j) = f_l(\overline{x_j}') \\
f'_l(x_j) = f'_l(\overline{x_j}) & f'_l(x'_j) = f'_l(\overline{x_j}') \\
\end{array}
$\\

The concrete definitions of these functions are as follows for $1\leq l < k$: \\

$\begin{array}{ll}
f_{l}(x_l) = \{0\} & f_{l}(x'_l) = \{0\} \\
f'_{l}(x_l) = \{1\} & f'_{l}(x'_l) = \{1\} \\
f_{l}(x_{l+1}) = \{0,1\} & f_{l}(x'_{l+1}) = \{-1\} \\
f'_{l}(x_{l+1}) = \{-1\} & f'_{l}(x'_{l+1}) = \{0,1\} \\
f_l(x_j) = f'_l(x'_j) = \{0,1\} & \forall j\notin\{l,l+1\} 
\\
\end{array}
$\\

Finally, we have an additional function $f_0$ to deal with the anchor case:\\

$
\begin{array}{l}
f_0(x_1) = f_0(\overline{x_1}) = f_0(x'_1) = f_0(\overline{x_1}') =\{-1\}
\\
f_0(x_j) = f_0(\overline{x_j}) = f_0(x'_j) = f_0(\overline{x_j}') =\{0,1\}\\\hspace{1em} ~~\text{for}~~  1< j \leq k
\end{array}$\\

\end{itemize}

Clearly, this transformation can be done in polynomial time
with respect to the size of the $\TQBF$ instance: being $n$ the number of clauses in $\phi$, there are $2k + n$ functions in the $\ND_1$ instance, $4k$ inputs, and $3$ outputs.
We prove that the transformation actually
reduces $\TQBF$ into $\ND_1$ in the appendix of the paper.

\eprfsketch

\section{Conclusions and future work}\label{sec:conclusions}

In this paper we have studied the computational complexity of adaptive testing in a setting where the set of possible definitions of the IUT and the behavior of each of them for each possible interaction are extensionally defined case by case. Problem variants where non-determinism is allowed or forbidden, and either the set of possible correct IUT definitions or the set of  incorrect ones must be a singleton or not, have been studied. Due to the propagation of complexity hardness by generalization, our complexity hardness results apply not only to our simple model, but also to any other more expressive adaptive testing models generalizing ours. In particular, the PSPACE-completeness of our two non-deterministic problem variants shows that states are not needed at all to make adaptive testing a PSPACE-hard problem, as a fully memory-less setting is assumed in them.

Regarding practical application, let us remark that in many practical scenarios it is very common to list the errors that could occur. Although this list can be relatively long when all of them are listed one by one, it is also very common that for practical purposes it can be described in a reasonably compact way. A good example of this situation is the one that can be seen in the example in Appendix A, 
where the repertoire of possible errors of the implementation under test is relatively short for a tester to express. However, this specification is easily translated into an enormously long list of possible implementations. Thus, although the tester does not need to name them one by one, it is easy to obtain such a detailed enumeration.

Our main line of future work, already ongoing work, consists in heuristically solving random and designed instances of our adaptive testing problems by applying heuristic, non-optimal methods. We are developing a minimax implementation to handle the PSPACE-complete variants, i.e., the non-deterministic ones, and an implementation based on Genetic Algorithms to handle the deterministic ones. Preliminary observations suggest that these methods could provide reasonable sub-optimal testing strategies in reasonable execution times.

\bigskip
\noindent\textbf{Acknowledgements:} 
Work partially supported by project PID2019-108528RB-C22, and by Comunidad de Madrid as part of the program S2018/TCS-4339 (BLOQUES-CM) co-funded by EIE Funds of the European Union. The authors would like to thank the anonymous reviewers for valuable suggestions on a previous version of this paper.

\section*{Appendixes}

\subsection*{Appendix A. Second motivational example: how to deal with larger scenarios}

In the example shown in Section~\ref{sec:example}, the number of possible IUT definitions (functions) to be considered was quite reduced, so the set of them was easy to enumerate manually. In more complex and realistic testing scenarios,
a big number of possible IUT definitions could emerge by combining, in any way deemed possible by the tester, all considered possible correct choices and all considered possible mistakes. Note that, given the lack of mandatory specification model ---e.g. FSMs, Timed Automata, C++ programs, etc.--- in our extensional setting, the tester has full freedom to consider in each scenario which mistakes can happen simultaneously and which ones cannot.
Let us point out that the existence of a huge set of possible definitions of the IUT does not imply a lengthy  definition {\it work} by the tester. In practice, those huge sets should usually be the result of explosively combining a much more manageable number of factors describing, in a compact fashion, all possible allowed and forbidden behaviors. In these cases, a simple program can be written to enumerate all the resulting functions, i.e. possible IUT definitions, by just exhaustively combining all factors in all ways considered possible. This way, the tester is released from the enumeration task. Next we present an example involving a big number of functions.

Let us assume that we test an already assembled automated teller machine (ATM). Note that, in this case, the definition of each test does not only consist in a way to interact with the ATM's keyboard, because the initial configuration of the ATM also affects the outcome of the test. This configuration involves the availability of notes of each type in the ATM and the state, in the bank database, of all user accounts connecting to ATM during the test.
Thus, the application of each test case of the ATM involves putting some exact amount of notes of each kind (10\euro, 20\euro, 50\euro, and 100\euro) into the cashier's banknote deposit, changing the database to create fake users with the desired balance amounts, interacting with the ATM via its keyboard, letting the money out, and resetting the system again. The whole process may take several minutes,
so it is not feasible to apply a huge number of test cases. Hence, our testing strategy must be designed so as to provide as much information about the IUT (in-)correctness as possible ---with a limited total interaction with it.

In this setting, a full test case, i.e. {\it input} in our setting, is defined in terms of the following data:

\begin{itemize}
    \item[A] Number of bills of each amount in the ATM before starting.
    \item[B] For each user ID involved in the withdrawals to be made during the test case, money in the account of that user before starting the test.
    \item[C] For each consecutive money withdrawal made during the same test,
    we have the following information: ID of the user making the withdrawal;
    amount of money being requested; and confirmation "yes / no" that will be answered in case the system requests it.
\end{itemize}

Information (A) can be denoted by a 4-tuple, using one element for each note type, (B) by a set of pairs (ID and money in account), and (C) by an ordered list of tuples. Hence, a test case can be denoted by a triple containing all three. For instance,
$$\begin{array}{ll}(&(24,13,14,23),\\
&\{(ID_1,100),(ID_2,140)\},\\
&[(ID_1,80,yes),(ID_2,200,no),(ID_1,40,yes)]\;)\end{array}$$
\noindent denotes a test where we initially provide the ATM with 24 10\euro\ bills, 13 20\euro\ bills, 14 50\euro\ bills, and 23 100\euro\ bills, next we create fake user accounts $ID_1$ and $ID_2$ with 100\euro\ and 140\euro, respectively, and finally we interact with the ATM making three consecutive withdrawal attempts involving these two users: $ID_1$ tries to withdraw $80$\euro, next $ID_2$ tries to take $200$\euro, and finally $ID_1$ comes back to try to withdraw $40$\euro.

Regarding the output of each test, it includes the following information for each withdrawal being requested: money given to the user, in particular numbers of notes given of each kind; information on the screen of the resulting balance of that user, which could be different to the actual balance being registered internally, though the latter is unknown to the user; and whether a confirmation ``yes / no'' is requested to the user or not. Moreover, we assume we can detect every time the ATM tries to give a bill that must exist in its notes deposit according to its internal data but in fact does not, so these errors are also included in the information associated with each withdrawal. Consequently, the output of each test will be represented by a list, where each element denotes all the previous output information for a withdrawal.

Regarding the specification, i.e. the set of possible IUT definitions being considered correct, some freedom is given to the implementer on several issues. We consider the following options:

\begin{itemize}
    \item (2 options) Return every $100$\euro\ to be delivered using a single $100$\euro\ bill; or two bills of $50$\euro\ each. If one of these types of bill runs out, then the rest of the money is provided by using the other type of bill.
    \item (2 options) Return the remainder of dividing the amount to be given by $100$ by using the maximum possible number of $20$\euro\ bills available at the ATM and then complete with $10$\euro\ bills; or with the maximum possible number of $10$\euro\ bills and then complete with $20$\euro\ bills if we run out of $10$\euro\ bills.
    \item (2 options) If there is not enough physical money deposited in the ATM, then the implementation can give everything that is left in the ATM; or not give anything.
    \item (3 options) If the client does not have enough money in his/her account, then the implementation can give him/her exactly what is left in the account; give nothing; or give on credit. In the latter case, a request for confirmation will be required.
    \item (2 options) Request confirmation in all cases concerned in the previous two items where some money could be given;   or not.
\item (2 options) Before giving the requested money or displaying that the operation cannot be done, show an ad of the bank's investment funds; or show an ad of bank's loans.
\end{itemize}

For each of the previous six choices, each correct implementation will be allowed to deterministically follow one of the available options. Moreover, in the last choice, it will also be allowed to follow an additional option: to non-deterministically choose any of both options in each case. Thus, this choice unleashes 3 possibilities instead of 2. By combining the resulting options in any possible way, we have $2 * 2 * 2 * 3 * 2 * 3 = 144$ possible correct functions, i.e. correct possible IUT definitions, in our specification set.

Regarding the set of possible incorrect functions, we consider the following possible errors:
\begin{itemize}
    \item[a] In the second and all subsequent money withdrawals during the same test,  assume the data of the first user ID regardless of whether the test tries to use different users.
    \item[b] Not updating the amount of bills that are available in the ATM after each withdrawal.
    \item[c] Not updating the amount of money available on the user's account after each withdrawal.
    \item[d] (2 options) Give one more or one less bill of the biggest type of bill which should be delivered.
    \item[e] (2 options) Give the part of the money amount being multiple of $100$ in alternative bills of $100$\euro\ and $50$\euro\ until availability, ignoring the two allowed criteria; or totally ignoring that part of the money, i.e. not giving it.
    \item[f] (2 options) Give the remaining part of the money amount, i.e. the remains of $100$, in alternative bills of $20$\euro\ and $10$\euro\ until availability, ignoring the two allowed criteria; or totally ignore that part of the money, i.e. not giving it.
    \item[g] (6 options) Believe that bills of any type $x \in \{10,20,50,100\}$ are interchangeable with bills of type $y \in \{10,20,50,100\}$ with $x \neq y$, both for delivering them and for accounting the number of each available in the ATM. Since this creates an equality relation between bills of type $x$ and those of type $y$ with $x\neq y$, and the order between $x$ and $y$ is irrelevant, 6 options arise.
    \item[h] Giving all requested money regardless of the amount of money of the user in the bank.
    \item[i] Never showing any ad.
    \item[j] (2 options) In confirmation requests, internally swap yes and no answers; or consider that all answers are the same as that given in the first withdrawal.
\end{itemize}

This gives us $1 + 1 + 1 + 2 + 2 + 2 + 6 + 1 + 1 + 2 = 19$ types of possible errors. For each of the 144 correct specifications mentioned above, we consider multiple incorrect versions that include one or more of the above errors. The tester assumes the probability that the development team made more than $5$ mistakes is negligible, so there are $\binom{19}{5} + \binom{19}{4} + \binom{19}{3} + \binom{19}{2} + \binom{19}{1} + 1 = 11,628 + 3,876 + 969 + 171 + 19 + 1 = 16,664$ possible ways to introducing any number of mistakes from $0$ to $5$ in any of these $144 $ possible correct IUT definitions. Thus, there are $144 * 16,664 = 2,399,616$ possible definitions of the IUT, of which $144$ are correct.

Since the execution of each test case will  take several minutes, we are interested in minimizing the number of tests that we need to apply, and we can reduce that number if we use adaptive testing instead of preset testing.
Note that, even if the first tests do not exhibit any behavior which is faulty by its own, i.e.~before checking the consistency of their outputs with {\it other} test results, they can show us which other tests should be applied next to chase potential faults more efficiently. For instance,

\begin{itemize}
    \item if some test shows that the part of the money being multiple of 100 is given with 100\euro\ notes, then checking whether 50\euro\ notes are dealt with correctly requires putting an insufficient amount of 100\euro\ notes in subsequent tests, as otherwise they will not be used by the ATM. We have the opposite situation if that part of the money is given with 50\euro\ notes;
    \item we have a similar situation for the remainder of $100$ and $20$\euro$-10$\euro\ notes;
    \item if some test detects that the ATM chooses to deliver {\it all} money available in the ATM when the user tries to withdraw more money than available in the ATM (instead of rejecting the withdrawal), then checking whether the accounting of notes in the ATM works as expected requires less interactions: in some tests, the ATM will tell us exactly how many notes it {\it thinks} it has;
    \item if some test detects that confirmations are not requested, then there is no need to execute tests aimed at detecting faults in the confirmation process.

\end{itemize}

This illustrates that the order in which test cases are applied generally affects the number of test cases we need to apply to collect useful information about the IUT correctness/incorrectness: if the first test cases focus on deciding which other test cases would be useful afterwards, depending on the outputs collected by these first tests, then reaching correctness/incorrectness diagnoses will require less interactions in general. Hence, the order of test cases in a testing strategy, and in particular in each of its branches, is crucial.

As we mentioned before, a simple program can enumerate all possible correct and incorrect definitions of the IUT by combining all correct and incorrect alternatives in all ways allowed by the tester. In general, this method gives the tester full flexibility to reflect any knowledge she has on the way she knows or suspects the IUT was developed. For instance, it could be the case that faults X and Y are not expected to happen together but X and Z are; that fault X is not expected if the developers choose legitimate choice W; etc.

In the resulting enumeration, each possible IUT definition will be a function associating inputs, i.e. test cases, and outputs. Actually, the behavior of functions needs to be defined only for some finite set of test cases considered significant and representative by the tester, out of which a suitable testing strategy will be designed. This finite set of test cases, having in general more elements than the amount of test cases we can afford to apply to the IUT, will be extracted from the set of all possible test cases, and each test case in the set will have the possibility to be included or not in the testing strategy.

In general, each of these {\it a priori} significant test cases will compose values from several finite, discrete, and/or continuous domains of choices depending on the nature of the IUT.
Examples of these possibilities include, respectively, simple choices such as 'yes'/'no', amounts of euros to be withdrawn such as 70\euro\ or 450\euro, or time lapses between consecutive keystrokes such as $3.47$ seconds or $0.1242$ seconds.\footnote{These time lapses are not considered in test cases in our example because no functionality depends on them. If a functionality such as e.g. logging out when a timeout is reached were added, then explicitly references to time lapses between consecutive interactions in the ATM would be needed in test cases.}
Note that, for choices involving continuous or infinite discrete domains, some representative elements have to be selected, out of infinite possibilities, to be included in the test cases under consideration.
Both extreme and standard values should be taken, where the former consist in very small and very big values, and the latter can be extracted by  picking several values with the same expected distribution as the one the IUT is expected to face in real life usage.
For instance, if some real value is expected to be distributed according to an exponential variable with $\lambda = 3$, then several samples of that variable could be used in the composition of the test cases under consideration, as well as some very small and very big values for the sake of degenerate behavior observation.

Notice that we do not need to
store a {\it program} to represent each function: a simple vector of legitimate and faulty choices can denote each function, and a single generic program receiving these choices will tell us the behavior of the corresponding function denoted by that vector.

Outputs do not denote continuous values in our example because no system requirement depends on any continuous domain in this case. Alternatively, let us consider some scenario where they do,
so we need to deal with outputs such as
e.g. $o = $ ``{\it signal Ok is shown after 3.448 seconds}.''
In this case, some error margin should be allowed when matching observed IUT outputs during testing. Thus, output $o$ should be redefined as e.g. ``{\it signal Ok is shown after 3.448 $\pm$ 0.05 seconds}''.
Let us note that these intervals could overlap and produce some non-determinism at the {\it observation} level of the testing process.
For example, let $f\in E$  and $f'\in C\backslash E$ with $f(a)=\{1.4\}$ and $f'(a)=\{1.5\}$.
That is, according to the possible correct definition of the IUT denoted by $f$, the numeric output $1.4$ is deterministically answered to input $a$, and according to possible incorrect definition $f'$, output $1.5$ is deterministically replied to $a$.
Let the observation precision margin of the tester be $\pm 0.2$.
Let us suppose the actual definition of the IUT is $f$. Note that, upon the reception of input $a$, the $1.4$ reply of the IUT will be seen by the tester as any value within $[1.2,1.6]$.
On the other hand, if the actual IUT definition is $f'$, then the $1.5$ answer will be seen as any value within $[1.3,1.7]$.
Note that the sub-interval $[1.3,1.6]$ is common to both functions, whereas any value within intervals $[1.2,1.3)$ and $(1.6,1.7]$ would uniquely identify the function producing that answer as either $f$ or $f'$, respectively.
In order to integrate this  observability limitation, the behavior of $f$ and $f'$ for input $a$ is redefined as follows: $f(a)=\{o_1,o_2\}$ and $f'(a)=\{o_2,o_3\}$, where these three new outputs $o_1,o_2,o_3$ abstractly represent the disjoint numeric intervals $[1.2,1.3)$, $[1.3,1.6]$, and $(1.6,1.7]$, respectively.

Thus, although the original definitions of $f$ and $f'$ produce deterministic responses $1.4$ and $1.5$ to input~$a$, respectively, after the redefinition their response is non-deterministic to reflect our limited capability to distinguish some observations from actual outputs.

Although the goal of this paper is just showing the computational complexity of, given the previous data, finding testing strategies, next we will briefly outline how these problems could be heuristically faced in practice. Testing strategies could be dynamically constructed similarly as strategies for other PSPACE-complete games are computationally found in run time as long as the game develops. After the result of each test case is collected, and thus some possible correct and incorrect IUT definitions are discarded, a minimax algorithm should be applied up to some limited depth ---due to the exponential size of the search space--- and the test case returned by the algorithm as the best first move to be made should be applied next to the IUT. Most interesting branches could be explored more deeply, and some heuristic should be applied in general at nodes at that maximum depth. For instance, the ratio between possible correct definitions and possible incorrect definitions could be used as heuristic in these nodes.

Actually, in complex testing scenarios, the realistic goal should not be guaranteeing that, in any situation, {\it all} possible correct or {\it all} possible incorrect IUT definitions will be discarded. On the contrary, it should be reaching configurations where the ratio between possible correct definitions and possible incorrect definitions is extreme in either way ---that is, even though the minority side is not fully discarded, its relative probability is very low. Thus, the actual goal would not be reaching the
exhaustiveness and the precise correctness/incorrectness diagnoses it allows to give, but just reaching some kind of pseudo-exhaustiveness.

Given the non-negligible time cost of executing a minimax algorithm to decide each new test case to be applied to the IUT (depending on the depth it aims to reach), it is essential to adjust that depth in such a way its execution takes a relatively short time ---in particular, in comparison with the time needed to {\it apply} each test case to the IUT. Otherwise, the testing process would be halted for a non-negligible time, waiting for the decision of which test case should be applied next, and a significant amount of time would be wasted. Note that, in this case, rather than executing the algorithm and waiting for it, that time could actually be used to apply a big amount of {\it any} test cases: even if these test cases are chosen without any algorithm-driven criterion, e.g. randomly, applying {\it many more} test cases could be better than applying many fewer thoroughly-designed test cases. Let us point out that the execution of the algorithm and the application of additional tests to the IUT can be trivially done {\it in parallel}, so the best of both choices can be trivially combined:
we can apply both those clever test cases selected by the algorithm {\it and also} other randomly designed test cases, in such a way that the testing process {\it never} has to halt until the next (good) test case to be applied is decided by the minimax algorithm. Moreover, note that the execution of the minimax algorithm could give us not only the best test case to be applied next, but also other interesting test cases to be applied next --other moves reaching nodes with good maximin values--, so at least part of these other test cases to be applied could not be that random after all.

This combination of best test cases (best first moves) and other tests is unnecessary when the application of each test case is expected to take several minutes, as the example developed before. In this case, the execution of the minimax algorithm up to a decent depth would take a significantly shorter time than the application of any test case to the IUT, so there would be no ``idle testing time'' to be filled with those test cases commented above: {\it all} tests applied to the IUT would be exactly the best moves discovered by the minimax algorithm.

\subsection*{Appendix B.
Proof of Theorem~\ref{th:NonDet}} \label{ap:proof}

\subsubsection*{B.1 Algorithm solving $\ND$ in polynomial space}
\label{ap:algoritmo}
The algorithm solving $\ND$ in polynomial space we informally described earlier is the following:

$l=1$; $h=|I|$; $k=min(k,h)$

{[Step }0{]}

$i[l]=\min(I)$

{[Step }1{]}

$o[l]=\min(\underset{f\in\left\{ f\in C:\,o[m]\in f(i[m])\,\forall m<l\right\} }{\bigcup}f(i[l]))$

{[Step }2{]}
\begin{itemize}
\item if $l<k$:
\begin{itemize}
\item $l = l + 1$; go to  {[Step }0{]}.
\end{itemize}
\item if $l=k$:
\begin{itemize}
\item if $\{f\in C:\,o[m]\in f(i[m])\,\,\forall m\leq k\}$ is a subset of either $C\backslash E$ or $E$, that is, in case there does not exist $f\in E$ such that $o[m]\in f(i[m])$ for all $1\leq m\leq k$ or there does not exist $f\in C\backslash E$ such that $o[m]\in f(i[m])$ for all $1\leq m\leq k$:
\begin{description}
\item [{while}]  ($l>0$ and \\ $o[l]=\max{\underset{f\in\left\{ f\in C:\,o[m]\in f(i[m])\,\forall m<l\right\} }{\bigcup}f(i[l])}$)\\
 $l = l - 1$
\item [{if}] $l=0$: \textbf{return YES}.
\item [{else}]
$o[l]$ is assigned to the next output and we go to {[Step }2{]}.
\end{description}
\item if $\{f\in C:\,o[m]\in f(i[m])\,\,\forall m\leq k\}$ is neither a subset of $C\backslash E$ nor a subset of $E$:
\begin{description}
\item [{while}] ($l>0$ and \\$i[l]$ is the maximum input $a\in I$ with \\ $a\neq i[m]\;\forall\;1\leq m\leq l-1$)\\
$l = l - 1$.
\item [{if}] $l=0$: \textbf{return NO}.
\item [{else}]
$i[l]$ is assigned to the next input $a\in I$ with $a\neq i[m]\;\forall\;1\leq m\leq l-1$ and we go to  {[Step }1{]}.
\end{description}
\end{itemize}
\end{itemize}

Obviously, in order to solve the optimization problem rather than the decision problem, we could iteratively
try $k=1$, then $k=2$, etc., until we find the first value returning YES.

The efficiency of the previous algorithm could be improved. For instance, the algorithm checks whether all functions belong to $E$ or all belong to $C\backslash E$ only at level $min(k,h)$, but this could also hold earlier, and extending the current branch after it holds is unnecessary.
Still, the algorithm runs  within polynomial memory as it is. In particular, as we noted earlier, $i$, $o$, and $l$ define the exploration point through the tree, and they can be defined with polynomial memory.

\subsubsection*{B.2 $\ND_1 \in \PSPACEh$}
\label{ap:demo}
Given the reduction presented in the proof sketch of Theorem~\ref{th:NonDet}, we have to prove that any instance
of $\TQBF$ is satisfied iff the corresponding instance of $\ND_1$ is satisfied.

\paragraph{B.2.1 $\TQBF \Rightarrow \ND_1$}

We start proving that in case the $\TQBF$ formula is satisfied, then the corresponding $\ND_1$ instance is also satisfied.

First, we prove that either $g$ or all the $f_l$ (and $f'_l$) functions will be discarded if the inputs corresponding to the existential variables are appropriately selected in the correct order, in particular as follows.
The first input can be any value
$i_{1}\in\{x_{1},\overline{x_{1}},x'_{1},\overline{x_{1}}'\}$.
Then, for the rest of inputs, we have to appropriately select them from the set $\{x_l,\overline{x_l}\}$ or from the set $\{x'_l,\overline{x_l}'\}$ to discard $f_l$ or $f'_l$, depending on the last output that was obtained. Formally, for each $1\leq l<k$:
\begin{itemize}
\item if $o_{l}=-1$ then the correct function $g$ is discarded and we know for sure that the IUT is incorrect. Thus,
no more inputs are needed at this strategy branch;
\item if $o_{l}=0$ (which does not discard $f_l$) then we use $i_{l+1}\in\{x'_{l+1},\overline{x_{l+1}}'\}$ (which does);
\item if $o_{l}=1$ (which does no discard $f'_l$) then we use $i_{l+1}\in\{x_{l+1},\overline{x_{l+1}}\}$ (which does).
\end{itemize}

Thus, in case any output is $-1$, then all the functions that are consistent with the inputs and outputs are incorrect, because the only correct function $g$ never returns $-1$, i.e., $g$ is discarded.
Otherwise, all the outputs belong to the set $\{0,1\}$,
but all the functions $f_l$ and $f'_l$ are discarded. Note that for all $1\leq l < k$ we have two cases for $o_l$:

\[
o_{l}=0\Rightarrow\begin{cases}
o_{l}\notin f'_{l}(i_{l})=\{1\}\\
i_{l+1}\in\{x'_{l+1},\overline{x_{l+1}}'\}\Rightarrow\\\hspace*{2em}0\leq o_{l+1}\notin f_{l}(i_{l+1})=\{-1\}
\end{cases}
\]
\[
o_{l}=1\Rightarrow\begin{cases}
o_{l}\notin f_{l}(i_{l})=\{0\}\\
i_{l+1}\in\{x_{l+1},\overline{x_{l+1}}\}\Rightarrow\\\hspace*{2em}0\leq o_{l+1}\notin f'_{l}(i_{l+1})=\{-1\}
\end{cases}
\]

Hence, in any case we will discard both $f_l$ and $f'_l$ for at least one input, which can be either $i_l$ or $i_{l+1}$. Finally, we have to discard also function $f_0$. However, taking into account that $i_{1}\in\{x_{1},\overline{x_{1}},x'_{1},\overline{x_{1}}'\}$
then we know that $0\leq o_{1}\notin f_{0}(i_{1})=\{-1\}$,
which discards $f_{0}$.

Thus, we conclude that for all $1\leq l < k$ we have $f_{l},f'_{l},f_{0}\notin\{f\in C:\,o_{m}\in f(i_{m}),\,m\leq k\}$.

Let us now prove that in case the qualified boolean formula is satisfied then either some output $-1$ is observed --so $g$ is discarded and the IUT is incorrect--, or
all functions $f_{c_{i}}$ will also be discarded.
In case variable $x_{1}$ is set to true in the valuation strategy applied to the $\TQBF$ instance, then we use $i_{1}\in\{x_{1},x'_{1}\}$. Otherwise, we use $i_{1}\in\{\overline{x_1},\overline{x_1}'\}$.
Then, for the rest of inputs, that is, for all
$1\leq l<k$ we have two options:
if $o_{l}<0$ then
no more inputs are needed,
because the correct function $g$ is discarded by $o_l$, as $g$ never returns $-1$; otherwise we take $y_l=o_l$ and have two cases again for the next input:

 $$\left\{\begin{array}{ll}
i_{l+1}\in\{x_{l+1},x'_{l+1}\} & \text{if }x_{l+1} \text{ is true in the }\\&\TQBF\text{ valuation}\\
i_{l+1}\in\{\overline{x_{l+1}},\overline{x_{l+1}}'\} & \text{if }x_{l+1} \text{ is false in the }\\&\TQBF\text{ valuation}
\end{array}\right.$$

That is, either all outputs are positive, or an output is negative and then we discard the correct function, so that all the functions that are consistent with the inputs and outputs are incorrect. Let us see that, in the latter case, all functions of the form $f_{c_{i}}$ are discarded.

Let us suppose that no observed output is negative. Then, for all disjunctive clauses $c_i$ of the logic formula $\phi$ and for all $1\leq j\leq k$,
\begin{itemize}
\item if $x_{j}$ is true then $i_{j}\in\{x_{j},\,x'_{j}\}$. If $x_{j}\in c_{i}$ then $0\leq o_{j}\notin f_{c_{i}}(i_{j})=\{-1\}$.
\item if $x_{j}$ is false then $i_{j}\in\{\overline{x_j},\overline{x_j}'\}$.
If $\overline{x_j}\in c_{i}$ then $0\leq o_{j}\notin f_{c_{i}}(i_{j})=\{-1\}$.
\item if $y_{j},\overline{y_j}\in c_{i}$ then $i_{j}\in\{x_{j},\overline{x_j},x'_{j},\overline{x_j}'\}. \text{ Thus, }0\leq o_{j}\notin f_{c_{i}}(i_{j})=\{-1\}$.
\item if $y_{j}\in c_{i}$ but $\overline{y_j}\notin c_{i}$ and $y_{j}$ is true, then
 $f_{c_{i}}(i_{j})$ is either $\{0\}$ or $\{-1\}$ for $i_{j}\in\{x_{j},\overline{x_j},x'_{j},\overline{x_j}'\}$
and $1=y_{j}=o_{j}\notin f_{c_{i}}(i_{j})\subset\{-1,0\}$.
\item if $\overline{y_j}\in c_{i}$ but $y_{j}\notin c_{i}$ and $y_{j}$ is false,
then $f_{c_{i}}(i_{j})$ is either $\{1\}$ or $\{-1\}$ for $i_{j}\in\{x_{j},\overline{x_j},x'_{j},\overline{x_j}'\}$
and $0=y_{j}=o_{j}\notin f_{c_{i}}(i_{l})\subset\{-1,1\}$.
\end{itemize}

All in all, if $c_{i}$ is satisfied, i.e.,~it has a literal whose value is true, then $f_{c_{i}}\notin\{f\in C:\,o_{m}\in f(i_{m}),\,m\leq k\}$. Hence, if $\phi$ is satisfied then all its clauses are satisfied and none of the  $f_{c_{i}}$ belongs to
$\{f\in C:\,o_{m}\in f(i_{m}),\,m\leq k\}$.

Let us remark that the election of inputs taken to discard functions $f_{c_i}$ is compatible with the election of inputs used to discard functions $f_{l}$, $f'_{l}$, and
$f_{0}$. In the second case, we choose between the inputs with and without primes, while in the first case we choose between the positive and negative inputs. By considering both factors together, a single input will be available in each situation for all inputs but the first one, where prime and non-prime versions are available. Therefore, by combining both strategies, in case we introduce the variables in the proper order and $\phi$ is satisfied, then we will discard either $g$ or all the other functions, i.e., those with forms $f_{c_{i}},f_{l},f_{0}$, and then we solve the $\ND_1$ problem.

\paragraph{B.2.2 $\ND_1 \Rightarrow \TQBF$}

Now we prove that the qualified boolean formula holds in case the corresponding instance of $\ND_1$ is satisfied.

First, we prove that if $(C,E,k)$ satisfies $\ND_1$ then, at each branch of the strategy, either there is some output $-1$ (and $g$ is discarded), or for all $1\leq j\leq k$ we have at least
one input from the set $\{x_{j},\overline{x_j},x'_{j},\overline{x_j}'\}$.
Thus, as we can only use $k$ inputs at any branch, we have exactly one
possible valuation for each $x_j$ in the latter kind of branches, i.e., those without $-1$.

Let us suppose $(C,E,k)$ satisfies $\ND_1$ and let us assume the inputs and outputs at the same branch are $\{i_{m}\}_{m=1}^{k}$ and
$\{o_{m}\}_{m=1}^{k}$. In case there exists $o_{m}=-1$ then
we have the property.
Thus, we have to look for contradictions in the other case. That is, let us now consider that the branch is such that, for all $1\leq m\leq k$, $o_{m}\in\{0,1\}$. Then, $g\in\left\{ f\in C:o_{m}\in f(i_{m}),i\leq k\right\} $. Moreover, as  $(C,E,k)$ satisfies $\ND_1$ we have that $E=\left\{ g\right\} =\left\{ f\in C:o_{m}\in f(i_{m}),i\leq k\right\} $, because
otherwise the tests would not allow to distinguish incorrect functions from the correct one as required to satisfy $\ND_1$. Let us prove that this property cannot hold if  $\exists j:\,x_{j},\overline{x_j},x'_{j},\overline{x_j}'\notin\{i_{m}\}_{m=1}^{k}$:

\begin{itemize}
\item If $j=1$ then $f_{0}$ could not be distinguished from $g$ with any other input, as 
$f_0$ is equal to $g$ for all inputs but those corresponding to $j=1$. Thus, $f_{0}\in\left\{ f\in C:o_{m}\in f(i_{m}),i\leq k\right\} $, and this is a contradiction, because the incorrect function $f_0$ would not be distinguished from the correct function $g$.
\item If $j>1$ and $\cancel{\exists}i_{M}\in\{x_{j-1},\overline{x_{j-1}},x'_{j-1},\overline{x_{j-1}}'\}$ then we have two consecutive variables which are not covered (those with subscripts $j$ and $j-1$), but functions $f_l$ only differ from $g$ for variables $l$ and $l+1$. That is, we will have that for all $m$: $f_{j-1}(i_{m})=f'_{j-1}(i_{m})=g(i_m)=\{0,1\}$. Thus $f_{j-1},f'_{j-1}\in\left\{ f\in C:o_{m}\in f(i_{m}),i\leq k\right\} $
and that is again a contradiction.
\item If $j>1$ and $\exists!i_{M}\in\{x_{j-1},\overline{x_{j-1}},x'_{j-1},\overline{x_{j-1}}'\}$
then $o_M$ can be either $0$ or $1$. In the first case,  $f_{j-1}\in\left\{ f\in C:o_{m}\in f(i_{m}),i\leq k\right\} $,
because $o_M$ does not discard $f_{j-1}$ and this function can only be discarded with inputs of the set
$\{x_{j-1},\overline{x_{j-1}},x'_{j-1},\overline{x_{j-1}}',
x_{j},\overline{x_j},x'_{j},\overline{x_j}'\}$, and we have 
not more inputs of that type. Analogously, in case $o_M$ is $1$ we have the same situation with $f'_{j-1}$. In any case, we have a contradiction, as either $f_{j-1}$ or $f'_{j-1}$ is not distinguished from $g$.
\item If $j>1$ and we have that $\exists i_{M_{1}},i_{M_{2}},\ldots,i_{M_{p}}\in\{x_{j-1},\overline{x_{j-1}},x'_{j-1},\overline{x_{j-1}}'\}$
then we are applying several inputs corresponding to variable $x_{j-1}$. However, for $f_{j-1}$ (respectively $f'_{j-1}$), all of them will return the value $0$ (respectively $1$), so we will be in the same situation as in the previous item.
That is, either $f_{j-1}$ or $f'_{j-1}$ will not be discarded.
\end{itemize}

Let us now prove that inputs are necessarily introduced in the same order as in the definition of the $\TQBF$ instance. That is, we have that if $(C,E,k)$ satisfies $\ND_1$ then $\forall j:\,i_{j}\in\{x_{j},\overline{x_j},x'_{j},\overline{x_j}'\}$.

Let us assume that $(C,E,k)$ satisfies $\ND_1$ and let us consider a strategy branch with set of inputs $\{i_{m}\}_{m=1}^{k}$ and the corresponding outputs
$\{o_{m}\}_{m=1}^{k}$. Since $(C,E,k)$ satisfies $\ND_1$, the functions of $C$
which are consistent with these inputs and outputs are either all correct or
all incorrect.

In case there exists an output whose value is $-1$, then we obviously have  $\left\{ f\in C:o_{m}\in f(i_{m}),i\leq k\right\} \subseteq C\backslash E$, because $E=\{g\}$ and $g$ only returns either $0$ or $1$. In the other case (that is, $o_{m}\in\{0,1\}$ for all $1\leq m\leq k$), we have
$g\in\left\{ f\in C:o_{m}\in f(i_{m}),i\leq k\right\} $.
Thus, as $(C,E,k)$ satisfies $\ND_1$, we must have
$E=\left\{ g\right\} =\left\{ f\in C:o_{m}\in f(i_{m}),i\leq k\right\} $. That is, we discard the rest of functions (all incorrect ones). Let us see that there is a contradiction if the inputs are introduced out of order.

Consider the minimum value $j$ such that
$i_{j}\notin\{x_{j},\overline{x_j},x'_{j},\overline{x_j}'\}$.
Then, we have that $i_{j}\in\{x_{n},\overline{x_n},x'_{n},\overline{x_n}'\}$
for some $n>j$.

Let us denote by $i_{l}$ the only input that belongs to $\{x_{n-1},\overline{x_{n-1}},x'_{n-1},\overline{x_{n-1}}'\}$ in $\{i_m\}_{m=1}^k$.
Then we have  $l>j$.
If $i_{j}\in\{x_{n},\overline{x_n}\}$ (respectively $i_{j}\in\{x'_{n},\overline{x_n}'\}$),
in case $o_{j}\in\{0,1\}$ we will have $f_{n-1}\; \text{(resp. \ensuremath{f'_{n-1}})}\in\left\{ f\in C:o_{m}\in f(i_{m}),i\leq\boldsymbol{j}\right\} $,
and in the branch of the strategy where $o_{l}=0$ (respectively $o_{l}=1$)
we will have $f_{n-1}$ (resp. $f'_{n-1}$)\;$\in\left\{ f\in C:o_{m}\in f(i_{m}),i\leq\boldsymbol{k}\right\} $,
as the only way to distinguish these functions from function $g$
is by using variables with subscripts $n$ or $n-1$. This makes the contradiction.

Now that we know that there is a single input for each variable and that the inputs are necessarily in the same order as the variables, that is,
 every input satisfies $i_{j}\in\{x_{j},\overline{x_j},x'_{j},\overline{x_j}'\}$,
we prove that, whenever  $(C,E,k)$ satisfies $\ND_1$, the corresponding formula $\exists x_1 \forall y_1 \ldots \exists x_k \forall y_k \;\phi$ is true, i.e., it satisfies $\TQBF$.
Let us show that, by following a $\TQBF$ strategy derived from the $\ND_1$ strategy used for instance $(C,E,k)$, every clause $c_i\in\phi$ is satisfied.
In order to satisfy $\phi$, in each $\TQBF$ strategy branch we set $x_{j}$ to true if $i_{j}\in\{x_{j},x'_{j}\}$ in the corresponding $\ND_1$ strategy branch, whereas we set it to false if $i_{j}\in\{\overline{x_j},\overline{x_j}'\}$.

Let us remind that the values of each $i_{j}$ (for $j>1$) depend
on the previous output $o_{j-1}$. In our case, we use $o_{j-1}=y_{j-1}$.
Note that $\ND_1$ strategy branches with some $-1$ output  trivially discard the correct function and make the testing strategy win in branches {\it not} existing in the $\TQBF$ strategy counterpart, where variables can only be true or false. As the $\TQBF$ strategy cannot include branches like this, next we focus on the other branches.
For all branches of the $\ND_1$ strategy where no $o_m$ is $-1$,
we know that for all $1\leq m\leq k$
we have $o_{m}\in\{0,1\}$. Hence,
$g\in\left\{ f\in C:o_{m}\in f(i_{m}),i\leq k\right\} $.
Therefore, as we know $(C,E,k)$ satisfies $\ND_1$,
we have that $E=\left\{ g\right\} =\left\{ f\in C:o_{m}\in f(i_{m}),i\leq k\right\} $ in those branches.

Under the previous conditions, let us show that if a function
$f_{c_{i}}$ is discarded, that is, we have that $f_{c_{i}}\notin\left\{ f\in C:o_{m}\in f(i_{m}),i\leq k\right\}$, then the corresponding clause
$c_i$ is satisfied.
In order to discard $f_{c_{i}}$, there must exist an input $i_{M}$ such that $o_{M}\notin f_{c_{i}}(i_{M})$.
Obviously, if $f_{c_i}(i_{M})=\{-1\}$ then the function is discarded, as we are considering branches where outputs $-1$ are not observed. By construction, function $f_{c_i}$ can be defined like this for input $i_M$ in three situations:
\begin{itemize}
\item $x_{M}\in c_{i}$ and $i_{M}\in\{x_{M},x'_{M}\}$, so $x_{M}$ is true.
\item $\overline{x_M}\in c_{i}$ and $i_{M}\in\{\overline{x_M},\overline{x_M}'\}$, so $x_{M}$ is false.
\item $y_{M},\overline{y_M}\in c_{i}$.
\end{itemize}
Obviously, in the three cases we have that clause $c_{i}$ is satisfied.

The rest of cases where $f_{c_i}$ is discarded are the following:
\begin{itemize}
\item if $y_{M}\in c_{i}$ but $x_{M},\overline{y_M}\notin c_{i}$ then
$f_{c_{i}}(i_{M})=\{0\}$ for $i_{M}\in\{x_{M},x'_{M}\}$. Thus, if
$f_{c_{i}}$ is discarded then $o_{M}=1$.
\item if $y_{M}\in c_{i}$ but $\overline{x_M},\overline{y_M}\notin c_{i}$ then $f_{c_{i}}(i_{M})=\{0\}$ for $i_{M}\in\{\overline{x_M},\overline{x_M}'\}$. Thus,
if $f_{c_{i}}$ is discarded then $o_{M}=1$.
\item if $\overline{y_M}\in c_{i}$ but $x_{M},y_{M}\notin c_{i}$ then
$f_{c_{i}}(i_{M})=\{1\}$ for $i_{M}\in\{x_{M},x'_{M}\}$. Thus,
if $f_{c_{i}}$ is discarded then $o_{M}=0$.
\item if $\overline{y_M}\in c_{i}$ but $\overline{x_M},y_{M}\notin c_{i}$
then $f_{c_{i}}(i_{M})=\{1\}$ for $i_{M}\in\{\overline{x_M},\overline{x_M}'\}$. Thus,
if $f_{c_{i}}$ is discarded then $o_{M}=0$.
\end{itemize}
As it can be seen, if $f_{c_{i}}$ is discarded then either $y_{M}\in c_{i}$ and $o_{M}=y_{M}=1$ or $\overline{y_M}\in c_{i}$ and $o_{M}=y_{M}=0$. In both cases, $c_{i}$ is satisfied.

In any other case,  $f_{c_{i}}$ is not discarded, because
in any other case $f_{c_{i}}(i_{M})=\{0,1\}$.

As we have seen that every $c_{i}$ is satisfied, we also have that  $\phi$ is satisfied.

\bibliographystyle{IEEEtran}

\bibliography{tse_bib}

@book{sipser12,
  title={Introduction to the Theory of Computation},
  author={Sipser, Michael},
  year={2012},
  publisher={Cengage Learning}
}

@article{stockmeyer87,
  title={Classifying the computational complexity of problems},
  author={Stockmeyer, Larry},
  journal={The journal of symbolic logic},
  volume={52},
  number={1},
  pages={1--43},
  year={1987},
  publisher={Cambridge University Press}
}

@article{feige1998threshold,
  title={A threshold of ln n for approximating set cover},
  author={Feige, Uriel},
  journal={Journal of the ACM (JACM)},
  volume={45},
  number={4},
  pages={634--652},
  year={1998},
  publisher={ACM New York, NY, USA}
}

@inproceedings{Crescenzi97,
  title={A short guide to approximation preserving reductions},
  author={Crescenzi, Pierluigi},
  booktitle={Proceedings of Computational Complexity. Twelfth Annual IEEE Conference},
  pages={262--273},
  year={1997},
  organization={IEEE}
}

@article{efatmaneshnik19,
  title={System Test Architecture Evaluation: A Probabilistic Modeling Approach},
  author={Efatmaneshnik, Mahmoud and Shoval, Shraga and Joiner, Keith},
  journal={IEEE Systems Journal},
  year={2019},
  volume={},
  number={},
  pages={1-12},
  publisher={IEEE}
}

@article{morell1990theory,
  title={A theory of fault-based testing},
  author={Morell, Larry J.},
  journal={IEEE Transactions on Software Engineering},
  volume={16},
  number={8},
  pages={844--857},
  year={1990},
  publisher={IEEE}
}

@inproceedings{ky15,
author = {Kushik, Natalia and Yevtushenko, Nina},
year = {2015},
month = {04},
pages = {73--78},
title = {Adaptive Homing is in {P}},
booktitle = {Tenth Workshop on Model-Based Testing (MBT 2015), EPTCS 180},
doi = {10.4204/EPTCS.180.5}
}

@article{yyk17,
  title={The complexity of checking the existence and derivation of adaptive synchronizing experiments for deterministic {FSMs}},
  author={Yenig{\"u}n, H{\"u}sn{\"u} and Yevtushenko, Nina and Kushik, Natalia},
  journal={Information Processing Letters},
  volume={127},
  pages={49--53},
  year={2017},
  publisher={Elsevier}
}

@inproceedings{kyy16,
  author    = {Natalia Kushik and
               Nina Yevtushenko and
               H{\"{u}}sn{\"{u}} Yenig{\"{u}}n},
  title     = "Reducing the Complexity of Checking the Existence and Derivation of
               Adaptive Synchronizing Experiments for Nondeterministic {FSMs}",
  booktitle = {International Workshop on domAin specific Model-based
               AppRoaches to vErificaTion and validaTiOn, AMARETTO@MODELSWARD 2016},
  pages     = {83--90},
  year      = {2016},
  publisher = {SciTePress}
}

@article{DBLP:journals/infsof/DorofeevaEMCY10,
  author    = {R. Dorofeeva and
               K. El-Fakih and
               S. Maag and
               A.R. Cavalli and
               N. Yevtushenko},
  title     = {{FSM}-based conformance testing methods: A survey annotated
               with experimental evaluation},
  journal   = {Information {\&} Software Technology},
  volume    = {52},
  number    = {12},
  year      = {2010},
  pages     = {1286-1297},
  ee        = {http://dx.doi.org/10.1016/j.infsof.2010.07.001},
  bibsource = {DBLP, http://dblp.uni-trier.de}
}

@inproceedings{sv03,
AUTHOR = "M. Stoelinga and F. Vaandrager",
TITLE = "A Testing Scenario for Probabilistic Automata",
BOOKTITLE = "30th Int. Colloquium on Automata, Languages and Programming, ICALP'03, LNCS 2719",
YEAR = "2003",
publisher="Springer",
PAGES= "464--477"}

@article{lnr06,
author="N. L\'opez and M. {N}{\'u}{\~n}ez and I. Rodr{\'\i}guez",
title="Specification, testing and implementation relations for symbolic-probabilistic systems",
journal="Theoretical Computer Science",
pages="228-248",
volume="353",
number="1--3",
year="2006"}

@article{sva01,
    author="J. Springintveld and F. Vaandrager
            and P.R. D{'A}rgenio",
    title ="Testing Timed Automata",
    journal ="Theoretical Computer Science",
    volume ="254",
    number ="1-2",
    pages = "225--257",
    year =  "2001",
    note="Previously appeared as Technical Report~CTIT-97-17, University of Twente, 1997"
}

@Article{ly96,
    author = "D. Lee and M. Yannakakis",
    title = "Principles and methods of testing finite state machines: A survey",
    journal ="Proceedings of the IEEE",
    year ="1996",
    volume ="84",
    number="8",
    pages = "1090--1123",
}

@Article{tre96,
    author = "J. Tretmans",
    title = "Conformance testing with labelled transition systems:
         Implementation relations and test generation",
    journal ="Computer Networks and ISDN Systems",
    year ="1996",
    volume ="29",
    pages = "49--79",
}

@inproceedings{tre99,
AUTHOR = "J. Tretmans",
title="Testing Concurrent Systems: A Formal Approach",
BOOKTITLE = "10th Int. Conf. on Concurrency Theory, CONCUR'99, LNCS 1664",
YEAR = "1999",
publisher="Springer",
PAGES= "46--65"}

@inproceedings{bt00,
  author = "E. Brinksma and J. Tretmans",
  title = "Testing Transition Systems: An Annotated Bibliography",
  booktitle = "4th Summer School on Modeling and Verification of Parallel Processes, MOVEP'00, LNCS 2067",
  pages="187--195",
  year = "2001",
  publisher="Springer"}

@inproceedings{pet00,
  author = "A. Petrenko",
  title = "Fault Model-Driven Test Derivation from Finite State Models: Annotated Bibliography",
  booktitle = "4th Summer School on Modeling and Verification of Parallel Processes, MOVEP'00, LNCS 2067",
  pages="196--205",
  year = "2001",
  publisher="Springer"}

@article{hie02b,
    author = "R.M. Hierons",
    title = "Comparing Test Sets and Criteria in the Presence of Test Hypotheses and Fault Domains",
    journal="ACM Trans. on Software Engineering and Methodology",
    volume="11",
    number="4",
    pages="427--448",
    year="2002"}

@inproceedings{kt04,
  author = "M. Krichen and S. Tripakis",
  title = "Black-box conformance testing for real-time systems",
  booktitle = "11th Int. SPIN Workshop on Model Checking of Software, SPIN'04, LNCS 2989",
  pages = "109--126",
  publisher = "Springer",
  year = "2004"
  }

@article{rmn08,
  author = "I. Rodr{\'\i}guez and M.G. Merayo and M. {N}{\'u}{\~n}ez",
  title = "{\HOTL}: Hypotheses and Observations Testing Logic",
  journal = "Journal of Logic and Algebraic Programming",
  volume = "74",
  number = "2",
  pages = "57--93",
  year = "2008"}

@inproceedings{bcf06,
AUTHOR = "I. Berrada and R. Castanet and P. F{\'e}lix and A. Salah",
title = "Test Case Minimization for Real-Time Systems Using Timed Bound Traces",
BOOKTITLE = "18th IFIP TC6/WG6.1 International Conference, TestCom 2006, LNCS 3964",
YEAR = "2006",
publisher="Springer",
PAGES= "289--305"}

@Article{mrr08,
  author =   "M.G. Merayo and M. {N}{\'u}{\~n}ez and I. Rodr{\'\i}guez",
  title =    "Extending {EFSMs} to specify and test timed systems with action durations and timeouts",
  journal =  "IEEE Transactions on Computers",
  volume =   "57",
  number =   "6",
  pages =    "835--844",
  year =     "2008",
}

@Article{hie09,
  author =   "R.M. Hierons",
  title =    "Verdict Functions in Testing with a Fault Domain or Test Hypotheses",
  journal =  "ACM Transactions on Software Engineering and Methodology",
  volume =   "18",
  number =    "4",
  year =     "2009",
}

@inproceedings{rod09c,
author="I. Rodr{\'\i}guez",
title="A general testability theory",
booktitle="CONCUR 2009 - Concurrency Theory, 20th International Conference, LNCS 5710",
pages="572--586",
publisher="Springer",
year="2009"
}

@ARTICLE{rlr14,
   author ="I. Rodr{\'\i}guez and L. Llana and P. Rabanal",
   title ="A general testability theory: classes, properties, complexity, and testing reductions",
   journal ="IEEE Transactions on Software Engineering",
   year ="2014",
   volume = "40",
   issue = "9",
   pages ="862--894"
}

@ARTICLE{rrr20,
   author = "I. Rodr{\'\i}guez and F. Rosa and F. Rubio",
   title = "Introducing complexity to formal testing",
   journal = "Journal of Logical and Algebraic Methods in Programming",
   year = "2020",
   volume = "111",
   issue = "Febraury"
}

@inproceedings{py14,
  author    = {Alexandre Petrenko and
               Nina Yevtushenko},
  title     = {Adaptive Testing of Nondeterministic Systems with {FSM}},
  booktitle = {15th International {IEEE} Symposium on High-Assurance Systems Engineering,
               {HASE} 2014},
  pages     = {224--228},
  year      = {2014}
}

@inproceedings{py11,
  author    = {Alexandre Petrenko and
               Nina Yevtushenko},
  title     = {Adaptive Testing of Deterministic Implementations Specified by Nondeterministic
               {FSMs}},
  booktitle = {Testing Software and Systems - 23rd {IFIP} {WG} 6.1 International
               Conference, {ICTSS} 2011},
  pages     = {162--178},
  year      = {2011}
}

@article{bv19,
  author    = {Petra van den Bos and
               Frits W. Vaandrager},
  title     = {State Identification for Labeled Transition Systems with Inputs and
               Outputs},
  journal   = {CoRR},
  volume    = {abs/1907.11034},
  year      = {2019},
  url       = {http://arxiv.org/abs/1907.11034},
  eprinttype = {arXiv},
  eprint    = {1907.11034},
  bibsource = {dblp computer science bibliography, https://dblp.org}
}

@article{bfg19,
  author    = {Roderick Bloem and Goerschwin Fey and Fabian Greif and Robert K\"onighofer and Ingo Pill and Heinz Riener and Franz R\"ock},
  title     = {Synthesizing adaptive test strategies from temporal logic specifications},
  journal   = {Formal Methods in System Design},
  volume    = {55},
  issue     = {2},
  pages     = {103--135},
  year      = {2019}
}

\end{document}